\pdfoutput=1

\documentclass[11pt,twoside,a4paper,cmspaper,final,collab]{cms-tdr}
\def\svnVersion{220878}\def\cmsCernNo{2013-195}\def\cmsCernDate{\today}\def\cmsMessage{Published in the European Physical Journal C as \href{http://dx.doi.org/10.1140/epjc/s10052-013-2674-5}{\doi{10.1140/epjc/s10052-013-2674-5.}}}

\begin{document}\cmsNoteHeader{FSQ-12-022}

\hyphenation{had-ron-i-za-tion}
\hyphenation{cal-or-i-me-ter}
\hyphenation{de-vices}

\RCS$Revision: 216786 $
\RCS$HeadURL: svn+ssh://svn.cern.ch/reps/tdr2/papers/FSQ-12-022/trunk/FSQ-12-022.tex $
\RCS$Id: FSQ-12-022.tex 216786 2013-11-17 15:44:06Z mazarkin $

\providecommand{\avg}[1]{\ensuremath{\langle #1 \rangle}}
\newlength\cmsFigWidth
\ifthenelse{\boolean{cms@external}}{\setlength\cmsFigWidth{0.95\columnwidth}}{\setlength\cmsFigWidth{0.75\textwidth}}
\cmsNoteHeader{FSQ-12-022} 

\title{Jet and underlying event properties as a function of charged-particle multiplicity in proton-proton collisions at $\sqrt{s}= 7\TeV$}
\ifthenelse{\boolean{cms@external}}{\titlerunning{Jet and underlying event properties\ldots}}{}

\date{\today}

\abstract{
Characteristics of multi-particle production in proton-proton collisions at $\sqrt{s}=7$\TeV are
studied as a function of the charged-particle multiplicity, $N_\text{ch}$. The produced particles are separated
into two classes: those belonging to jets and those belonging to the underlying event.
Charged particles are measured with pseudorapidity $\abs{\eta}< 2.4$ and transverse momentum $\PT> 0.25\GeVc$.
Jets are reconstructed from charged-particles only and required to have $\PT> 5\GeVc$. The
distributions of jet $\PT$, average $\PT$ of charged particles belonging to the underlying event
or to jets, jet rates, and jet shapes are presented as functions of $N_\text{ch}$ and compared to the
predictions of the \PYTHIA and \HERWIG event generators. Predictions without multi-parton
interactions fail completely to describe the $N_\text{ch}$-dependence observed in the data. For increasing
$N_\text{ch}$, \PYTHIA systematically predicts higher jet rates and harder $\PT$ spectra than
seen in the data, whereas \HERWIG shows the opposite trends. At the highest multiplicity, the
data--model agreement is worse for most observables, indicating the need for further tuning and/or new model
ingredients.
}

\hypersetup{%
pdfauthor={CMS Collaboration},%
pdftitle={Jet and underlying event properties as a function of particle  multiplicity in proton-proton collisions
at sqrt(s) = 7 TeV},%
pdfsubject={CMS},%
pdfkeywords={CMS, physics, underlying event, multiplicity}}

\maketitle 

\section{Introduction}

Achieving a complete understanding of the details of multi-particle production in hadronic collisions remains
an open problem in high-energy particle physics. In proton-proton (pp) collisions at the energies of the Large Hadron
Collider (LHC), most of the inelastic particle production is described in a picture in which an event is a combination
of hadronic jets, originating from hard parton-parton interactions with exchanged momenta above several \GeVc,
and of an underlying event consisting of softer parton-parton interactions, and of proton remnants.

The production of high-transverse-momentum jets, defined as collimated bunches of hadrons, results from
parton cascades generated by the scattered quarks and gluons, described by perturbative quantum
chromodynamics (QCD), followed by non-perturbative hadronization described either
via color fields (``strings") stretching between final partons, or by the formation of colorless clusters of hadrons~\cite{pQCDBook}.
The underlying event (UE) is commonly defined as the set of all final-state particles that are not associated
with the initial hard-parton scattering.
This component is presumably dominated by perturbative (mini)jets with relatively small
transverse momenta of a few\GeVc,  produced in
softer multi-parton interactions (MPI)~\cite{LUND, b11,ssz82,sj87,dnec2,dos12,bdfs12},
as well as by soft hadronic strings from the high-rapidity remnants. The description of the UE is more
phenomenological than that of the jets arising from the primary hard-parton scatter, whose final hadron
multiplicity can be in principle computed in QCD~\cite{pQCDBook}.
In this two-component approach, rare high-multiplicity
events can be explained as due to a large number of MPI taking place in the pp collisions at small impact
parameters. Different variants of such a physical picture are
realized in state-of-the-art Monte Carlo (MC) event generators such as \PYTHIA~\cite{PYTHIA6,PYTHIA8} and \HERWIG~\cite{HERWIGmain, HERWIG}.
The properties of multi-particle production are very sensitive to the assumptions made about the combination of MPI and hard scatterings,
the modeling of the multi-parton interactions (in particular the transverse structure of the
proton)~\cite{b11}, and non-perturbative final-state effects such as color reconnections, hadronization
mechanisms, and possible collective-flow phenomena, among others.

Experimental data on multi-particle production in pp collisions at LHC energies provide a
clear indication that our understanding of the different components contributing to the total inelastic cross
section is incomplete. This arises from difficulties in
describing multiplicity distributions, and especially the high-multiplicity tails~\cite{cmsmult}, or in
reproducing a new structure of the azimuthal angular correlations at 7\TeV for high-multiplicity events, the
so-called ``ridge''~\cite{cms}. Interesting disagreements between data and MC simulation were also recently reported in transverse
sphericity analyses and for global event shapes~\cite{AL, ATL, CMSevent}.
Together with similar findings in nucleus-nucleus collisions, these disagreements point to the intriguing possibility of some mechanisms
at high multiplicities which are not properly accounted for in event generator models.
Therefore, although the standard mixture of (semi)hard and non-perturbative physics considered by
\PYTHIA and \HERWIG is often sufficient for reproducing the bulk properties of inelastic events, it fails to provide
a more detailed description of the data and in particular of the properties of events binned in particle
multiplicity.

The average transverse momentum of the charged particles produced in pp and p{\=p} collisions has been
measured as a function of the event multiplicity at various center-of-mass energies
~\cite{cmsmult, Albajar:1989an, Acosta:2001rm, Aaltonen:2009ne,  Aad:2010ac, Aamodt:2010my}.
The work presented here is the first one that carries out the study also for the UE and jets separately and includes
other observables (jet $\PT$ spectra, rates and shapes) not analyzed before as a function of particle
multiplicity with such a level of detail.

The paper is organized as follows. The general procedure of the analysis is described in Section~\ref{analysis},
a short description of the Compact Muon Solenoid (CMS) detector is given in Section~\ref{detector}, and the event generator models
used are presented in Section~\ref{models}. Sections~\ref{evsel} to~\ref{syst} describe trigger and event
selection, track and jet reconstruction, the data correction procedure, and the systematic uncertainties.
Results and discussions are presented in Section~\ref{results}, and summarized in Section~\ref{conclusion}.

\section{Analysis strategy}
\label{analysis}

The main goal of this analysis is to study the characteristic features and relative importance of different mechanisms of
multi-particle production in pp collisions at a center-of-mass energy of $\sqrt{s}=7$\TeV in different charged-particle
multiplicity bins, corresponding to different levels of
hadronic activity resulting from larger or smaller transverse overlap of the colliding protons.
Guided by the two-component physical picture described in the introduction, we separate the particle content of each
inelastic event into two subsets. We identify the jet-induced contribution and treat the rest as the
underlying event originating from unresolved perturbative sources such as semihard MPI and other softer mechanisms.
Our approach to this problem uses the following procedure, applied at the stable (lifetime $c\tau>10\mm$) particle-level:
\begin{itemize}
 \item{Similarly to the centrality classification of events in high-energy nuclear collisions~\cite{hin},
     events are sorted according to their charged-particle multiplicity (Table~\ref{table_Nevt}). 
     Hereafter, for simplicity, multiplicity should always be understood as charged-particle multiplicity.}
\item{For each event, jets are built with charged particles only using the anti-$\kt$ algorithm~\cite{antikt, FastJet}
     with a distance parameter 0.5, optimized as described below, and are required to have a $\PT >5\GeVc$.
     Charged particles falling within a jet cone are labeled as ``intrajet particles". }
 \item{After removing all intrajet particles from the event, the remaining charged particles are defined as
    belonging to the underlying event. Events without jets above $\PT=5\GeVc$ are considered to consist of particles from the UE only. }
\end{itemize}

In order to achieve a better separation of the contributions due to jets and underlying event, the
resolution parameter of the  anti-$\kt$ algorithm is increased until the UE charged-particle $\PT$-spectrum starts to saturate,
indicating that the jet component has been effectively removed. This way of fixing the jet cone radius minimizes
contamination of the underlying event by jet contributions or vice versa. A resolution parameter of value 0.5 is
found to be optimal. Of course, it is not possible to completely avoid mixing between jets and underlying event.
To clarify the picture and minimize the mixing of the two components,  we measure not only the $\PT$ spectrum of
the charged particles inside jet cones, but also the spectrum of the leading (the highest-\PT) charged particle in each cone.

\section{The CMS detector}
\label{detector}

A detailed description of the CMS detector can be found in Ref.~\cite{cmsd}. A right-handed coordinate system with
the origin at the nominal interaction point (IP) is used, with the $x$ axis pointing to the center of the LHC
ring, the $y$ axis pointing up, and the $z$ axis oriented along the anticlockwise-beam direction. The
central feature of the CMS detector is a superconducting solenoid of 6\unit{m} internal diameter providing
an axial magnetic field with a nominal strength of 3.8\unit{T}. Immersed in the magnetic field are the pixel
tracker, the silicon-strip tracker, the lead tungstate electromagnetic calorimeter, the
brass/scintillator hadron calorimeter, and the muon detection system. In addition to the barrel
and endcap calorimeters, the steel/quartz-fibre forward calorimeter covers the pseudorapidity region
$2.9 <\abs{\eta}<5.2$, where $\eta = -\log\left[\tan\left(\theta/2\right)\right]$, and $\theta$ is the polar angle measured
at the center of the CMS detector with respect to the $z$ axis. The tracking detector consists of
1440 silicon-pixel and 15\,148 silicon-strip detector modules. The barrel part consists of 3 (10) layers of pixel (strip) modules around the IP
at distances ranging from 4.4\cm to  1.1\unit{m}. Five out of the ten strip layers
are double-sided and provide additional $z$ coordinate measurements. The two endcaps consist of 2 (12)
disks of pixel (strip) modules that extend the pseudorapidity acceptance to $\abs{\eta} = 2.5$. The tracker provides an
impact parameter resolution of about 100\micron and a $\PT$ resolution of about 0.7\% for 1\GeVc charged
particles at normal incidence.
Two of the CMS subdetectors acting as LHC beam monitors, the Beam Scintillation Counters (BSC) and the
Beam Pick-up Timing for the eXperiments (BPTX) devices, are used to trigger the detector readout. The BSC
are located along the beam line on each side of the IP at a distance of 10.86\unit{m} and cover  the range
$3.23 <\abs{\eta} < 4.65$. The two BPTX devices, which are located inside the beam pipe and $\pm$175\unit{m} from
the IP, are designed to provide precise information on the structure and timing of the LHC beams with a time resolution of 0.2\unit{ns}.

\section{Event generator models}\label{models}

The best available general-purpose event generators and their tunes  are used for comparison with the data.
They are  the \PYTHIA~6 (version 6.424~\cite{PYTHIA6}, tune Z2*), \PYTHIA~8 (version 8.145 ~\cite{PYTHIA8}, tune 4C~\cite{4C}),
and {\HERWIG}++ 2.5 (tune \textsc{UE-EE-3M})~\cite{HERWIG} event generators. These event generators and tunes differ
in the treatment of initial and final state radiation, hadronization, and in the choice of underlying event parameters, color reconnections, 
and cutoff values for the MPI mechanism. Values of these parameters were chosen to provide a reasonable
description of existing LHC pp differential data measured in minimum-bias and hard QCD processes.
Initial and final state radiation is essential for the correct description of jet production and of the UE~\cite{PartonRadiation}. 
For the MPI modeling, \PYTHIA incorporates interleaved evolution
between the different scatterings~\cite{NonPertEf, 4C}, whereas \HERWIG concentrates more hard scatterings at the center of the pp
collision while allowing for more (disconnected) soft-parton scatterings at the periphery. 
A detailed review of the implementation of all these mechanisms in modern MC event generators is given~\cite{AllGen}. 
The most recent \PYTHIA~6 Z2* tune is derived from the Z1 tune~\cite{PYTHIA6Z1}, which uses the CTEQ5L parton distribution
set, whereas  Z2* adopts CTEQ6L~\cite{Pumplin:2002vw}. The Z2* tune
is the result  of retuning the \PYTHIA parameters PARP(82) and PARP(90) by means of the automated \textsc{Professor} tool~\cite{ProfessorTool},
yielding PARP(82)=1.921 and PARP(90)=0.227. The  results of this study are also compared to predictions obtained with
\PYTHIA~8, tune 4C, with multi-parton interactions switched off.
Hadronization in \PYTHIA is based on the Lund string  model~\cite{LUND} while that in {\HERWIG} is based on the cluster fragmentation
picture in which perturbative evolution forms preconfined clusters that subsequently decay into final hadrons. The version of {\HERWIG}++~2.5~\textsc{UE-EE-3M}
used in this paper includes important final-state effects due to color reconnections and is based on the MRST2008 parton distribution set~\cite{MRST}.

\section{Event selection and reconstruction}
\label{evsel}

The present analysis uses the low-pileup data recorded during the first period of 2010 data taking,
corresponding to an integrated luminosity of $(3.18 \pm 0.14)\pbinv$.
The data are collected using a minimum-bias trigger requiring a signal from both BPTX detectors coincident with a
signal from both BSC detectors.

For this analysis, the position of the reconstructed primary vertex is
constrained to be within $\pm$10\cm with respect to the nominal IP along the beam direction and within $\pm$2\cm
in the transverse direction, thereby substantially rejecting non-collision events~\cite{TrackVert7TeV}. The
fraction of background events after these selections is found to be negligible ($<$0.1\%).

The fraction of events in the data sample with pileup (two or more pp collisions per bunch crossing) varies
in the range (0.4--7.8)\% depending on the instantaneous luminosity per bunch. This small fraction of pileup
events is kept, but the analysis is only carried out for the tracks connected with
the primary (highest multiplicity) vertex. The fraction of events where two event vertices are
reconstructed as one, or where two event vertices share associated tracks, ranges between (0.04--0.2)\%.

\subsection{ Track reconstruction and selection}
\label{tracksel}

The track reconstruction procedure uses information from both pixel and strip detectors and is based on an
iterative combinatorial track finder~\cite{TrkVtxReco}.
Tracks are selected for analysis if they have transverse momenta $\PT>0.25$\GeVc and pseudorapidities lying
within the tracker acceptance $\abs{\eta}<2.4$. Such $\PT$ cut provides robust measurements, keeping the
event selection minimally biased by hard processes. In addition, tracks must be associated with the event vertex
with the highest multiplicity in the bunch crossing. The requirement removes
tracks coming from secondary interactions with detector materials,  decays of long-lived neutral hadrons, and pileup.
Residual contamination from such tracks is at the level of 0.2\%.

\subsection{Charged-particle jet reconstruction}

This analysis is based on jets that are reconstructed using tracks only, in
order to avoid the reconstructed jet energy uncertainty due to mismeasurements of low-$\PT$ neutral particles.
Jets are reconstructed by clustering
the tracks with the collinear- and infrared-safe anti-$\kt$ algorithm with a distance parameter of 0.5, that
results in cone-shaped jets.
Jets are retained if their axes lie within the fiducial region $\abs{\eta^\text{jet axis}}<1.9$, so that for a
jet with an effective radius of 0.5 all jet constituent tracks fall within the tracker acceptance ($\abs{\eta}<2.4$).

\section{Data correction}
\label{corrections}

\subsection{Event selection efficiency}

In the MC simulations, events are selected at the stable-particle level (lifetime $c\tau >10\mm$) if at least
one charged particle is produced on each side of the interaction point within $3.32  < \abs{\eta} <  4.65$,
mimicking the BSC trigger requirement, and, in addition, if at least five charged particles with $\PT >$
0.25\GeVc and $\abs{\eta} < 2.4$ are present, which ensures a high vertex finding efficiency in the offline
selection of data.

The trigger efficiency is measured using data collected with a zero-bias trigger, constructed from a
coincidence of the BPTX counters, which effectively requires only the presence of colliding beams at the
interaction point.  The offline selection efficiency is determined from MC simulations.  The combined trigger
and offline selection efficiency is obtained as a function of the number of reconstructed tracks and is very
high: above 87\% for events with more than 10 reconstructed tracks and close to 100\% for events with more
than 30 reconstructed tracks.  Results are corrected by applying a weight inversely proportional to the
efficiency to each observed event.

\subsection{Corrections related to the track reconstruction}
\label{correction:Mult}

The track-based quantities ($N_\text{ch}$, average \PT of tracks, jet \PT density in ring zones) are corrected in a two-stage correction procedure.
First, each observed track is given a weight to account for track reconstruction inefficiencies and misreconstructed (fake) track rates,
as obtained from the detector simulation.  The weights are based on two-dimensional matrices  $\epsilon(\eta, \PT)$ and $f(\eta,\PT)$,
for reconstruction efficiency and fake track rates, respectively, computed in bins in $\eta$, \PT, and is given by

\begin{equation}
N^\text{true}_\text{ch}(\eta, \PT)=N^\text{reco}_\text{ch}(\eta, \PT)\frac{{1-f(\eta, \PT)}}{\epsilon(\eta, \PT)}.
\end{equation}

The corrections for reconstruction efficiencies and fake rates depend on track multiplicity.
Therefore, four different sets of matrices $\epsilon(\eta, \PT)$ and $f(\eta,\PT)$ for different track multiplicity classes
are used, the first three track multiplicity classes corresponding
to the first three charged-particle multiplicity bins of Table~\ref{table_Nevt} and the fourth one corresponding to the fourth and fifth
charged-particle multiplicity bins.  The average track reconstruction efficiency and fake rate vary between 79--80\% and 3--4\%,
respectively, depending on the multiplicity bin considered.

Table~\ref{table_Nevt} shows the corrected charged-particle multiplicity classes used in this analysis and the number of events and mean multiplicities in
each multiplicity bin after applying all event selection criteria.
\begin{table*}
\topcaption{Charged-particle multiplicity bins, mean charged-particle multiplicity in bins, and corresponding number of events.
The multiplicity $N_\text{ch}$ is defined as the total number of stable charged-particles in the events, corrected for inefficiencies, with transverse
momentum $\PT>0.25$\GeVc and pseudorapidity $\abs{\eta}  <  2.4$.} \label{table_Nevt}
\centering
\begin{tabular}{ccccc}\hline
  Multiplicity range & Mean multiplicity $\langle N_\text{ch} \rangle$ & Number of events \\ \hline
   $10 < N_\text{ch} \leq 30$ & 18.9 &  2 795 688\\
   $30 < N_\text{ch} \leq 50$ & 38.8  &  1 271 987\\
   $50 < N_\text{ch} \leq 80$ & 61.4 &  627 731 \\
   $80 < N_\text{ch} \leq 110$ & 90.6 &  105 660 \\
   $110 < N_\text{ch} \leq 140$ & 120 &  11 599 \\ \hline
\end{tabular}
\end{table*}

\begin{figure}
\begin{center}
\includegraphics[ width=\cmsFigWidth]{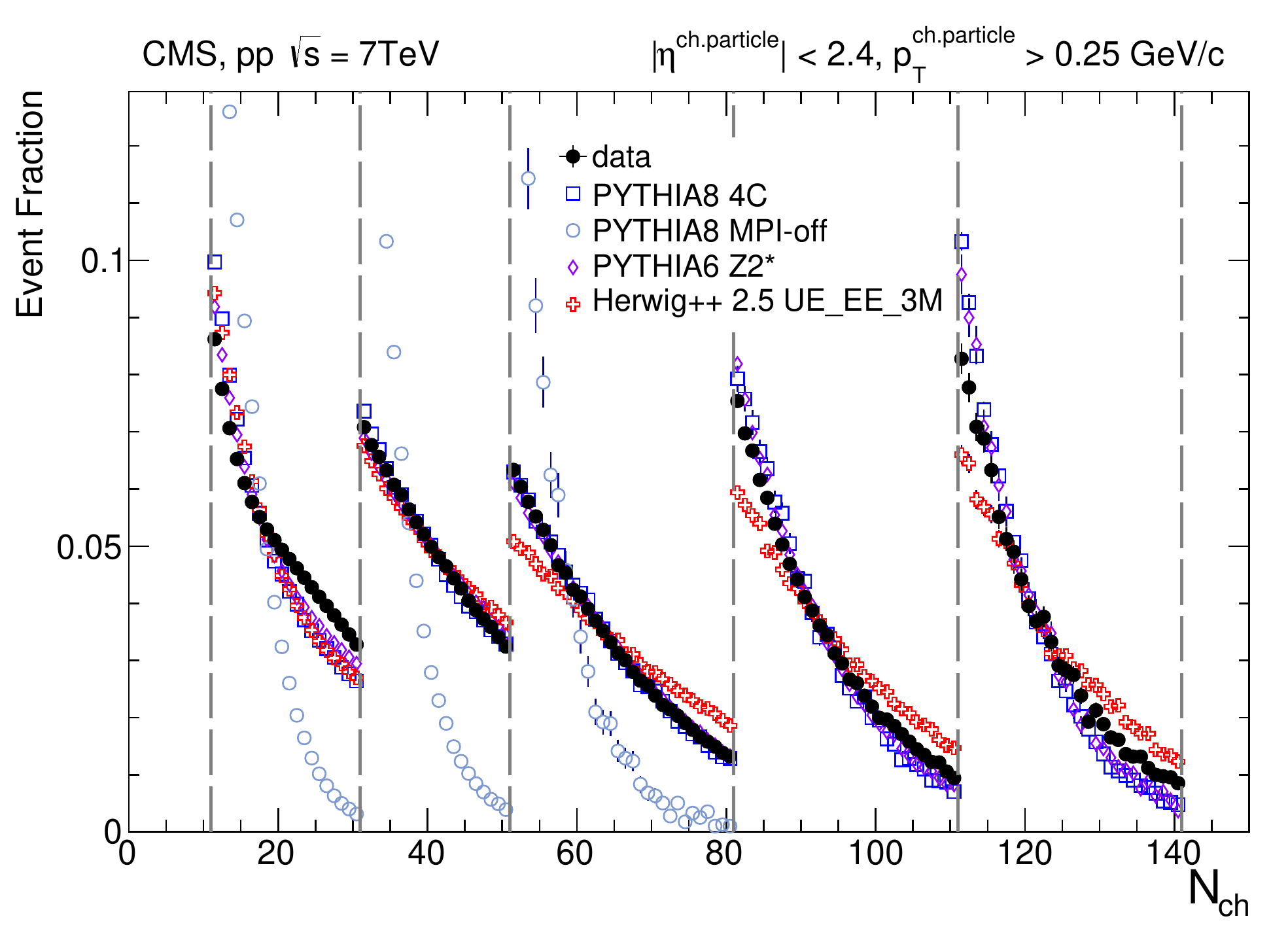}%
\caption{Charged-particle multiplicity distributions, corrected for tracking efficiency and fake rate, for the five multiplicity bins defined in this analysis
  compared to four different MC predictions. The normalization is done for each multiplicity bin separately.
  \PYTHIA~8 with MPI switched off completely fails to produce events at large multiplicity and therefore no points are shown in the two highest multiplicity domains.}%
\label{MultDistributions}%
\end{center}
\end{figure}

Figure~\ref{MultDistributions} shows multiplicity distributions that have been corrected for tracking efficiency and fake rate.
The simulations fail to describe all the measured $N_\text{ch}$ distributions, as discussed
in Ref.~\cite{cmsmult}.  As we are considering event properties as a function of multiplicity, such a data--MC disagreement might introduce a bias due to
the different $N_\text{ch}$ distribution within the wide multiplicity intervals.	
Reweighting the multiplicity distributions in MC simulations to bring them in agreement with the ones observed
in data  results in less than  1--2\% corrections for all results.
In the following, corrected results are compared to the predictions obtained from the unweighted MC models.

All the measured quantities hereafter are further corrected to stable-particle level using a bin-by-bin factor obtained from Monte Carlo simulations.
This correction factor accounts for event migration between adjacent multiplicity bins, for differences in the tracking performance in the dense environment
inside jets, and for mixing between charged particles belonging to charged-particle jets and the UE due to jets that are misidentified at the detector level.
The magnitude of this correction factor is typically less than 1\%, except for the jet $\PT$ density in the core of the jet where it reaches up to 8\%.

\subsection{Correction of the track-jet \texorpdfstring{\PT}{pt} distributions}

Track-jet distributions have to be corrected for inefficiencies in reconstruction, for misidentified jets, and for bin
migrations due to the finite energy resolution.
On average, a reconstructed track-jet has 95\% of the energy of the original charged-particle jet. The
energy resolution of such jets is about 13\%.
 The reconstructed jet spectrum is related to the ``true'' jet spectrum as follows:
\begin{equation}\label{resp}
M(\PT^\text{measured}) = \int C(\PT^\text{measured},\PT^\text{true})T(\PT^\text{true})\rd\PT^\text{true},
\end{equation}
where $M(\PT^\text{measured})$ and $T(\PT^\text{true})$ are the measured and the true $\PT$ spectra, respectively, and
$C(\PT^\text{measured},\PT^\text{true})$ is a response function obtained from the MC simulation.
The problem of inverting the response relation of Eq.~(\ref{resp}) is well known and has been extensively studied in literature.
In our analysis, an iterative unfolding technique~\cite{Unfolding} is applied. Since the detector response
changes with multiplicity, individual response matrices are used for each multiplicity bin.

\section{Systematic uncertainties}\label{syst}

The following sources of systematic uncertainties are considered:

\textbf{ Association of tracks with the  primary vertex (track selection): }
Tracks that are coming from a non-primary interaction result in an incorrect multiplicity classification of
the event and bias the event properties at a given multiplicity. These tracks originate from secondary interactions with
detector material, decays of long-lived neutral hadrons, and pileup. Moreover, these tracks can bias the $\PT$ spectrum of primary tracks.
As it is not possible to completely avoid contamination by such tracks,
the stability of the results has been estimated by tightening and loosening the association criteria. Removing
contamination inevitably leads to rejection of some valid primary tracks, so for each set of the association
criteria a special efficiency and fake-rate correction must be used.

\textbf{ Tracking performance: }
A correct description of the tracking performance in the MC simulation of the detector is essential.
A conservative estimate of the uncertainty of this efficiency of 2.3\% is taken from Ref.~\cite{Nch_1}.

\textbf{ Model dependence of the correction procedures: }
Different MC models can give slightly different detector and reconstruction responses.
Two models, \PYTHIA~6  tune Z2* and \PYTHIA~8 tune 4C, are used to compute tracking
and jet performance and correction factors. {\HERWIG}++~2.5 was found to deviate too much from the data and was
not used for the estimate of the systematic uncertainty. Corrections based on the \PYTHIA~6  tune Z2* model, which
provides better agreement with data, are used to get the central values of different physics quantities. The
differences between these two methods are assigned as a systematic uncertainty.

\textbf{ Unfolding the jet $\PT$ spectrum: }
The unfolding procedure used to correct for bin migrations in the jet $\PT$ spectra is based on an iterative
unfolding technique~\cite{Unfolding}  for which we find that 4--5 iterations are optimal. By varying the number of
iterations ($\pm$1 with respect to the optimal value) and the reconstructed-to-generated jet matching parameter ($0.15<
\Delta R <0.25$) we obtain a systematic uncertainty of $(0.5-2.0)\%$. This leads to a systematic uncertainty $<$0.2\%
in the average $\PT$ of the jet spectrum, and $<$2\% for charged-particle jet rates.

Although this analysis uses a low-pileup data sample, rare high-multiplicity events
might occur due to overlapping pp collisions in the same bunch crossing. The effect of pileup is
estimated by comparing results at different instantaneous luminosities. The dataset is divided
into subsets according to the instantaneous luminosity and
the differences found between these subsets are of the order of the statistical uncertainties of the sample.
In addition, it was checked that the instantaneous luminosity for events with small and large $N_\text{ch}$  does not differ,
confirming that the large-multiplicity bins are not biased by a possibly increased contribution from pileup events.
Therefore, we conclude that high-multiplicity events are not  affected by pileup.

Tables \ref{table:SumSysError} and \ref{table:SumSysError2} summarize the systematic and statistical
uncertainties of the measured quantities. The total uncertainties are the sum in quadrature of the individual
systematic and statistical uncertainties. The total error of  jet $\PT$ density as a function of jet radius
rises with R and $N_\text{ch}$.  The total uncertainties in the jet $\PT$ spectra are of the order of 4--8\%
for jet $\PT$ up to about 25\GeVc. For jets with  $\PT > 25\GeVc$ the statistical uncertainties dominate.

\begin{table*}
\topcaption{Summary of systematic and statistical uncertainties for various averaged inclusive and UE-related quantities.
The variables  $\avg{ \PT^\text{ch. particle}}$,  $\avg{\PT^\mathrm{UE}}$, $\avg{PT^\mathrm{ij}}$,
 $\avg{\PT^\mathrm{ijl}}$ are defined in Section~\ref{sec:GeneralProperties}, $\rho(R)$ is defined in Section~\ref{sec:JetWidth}.}
\label{table:SumSysError}
\centering
\begin{tabular}{lccccc}
\hline
  & $\avg{\PT^\text{ch. particle}}$ & $\avg{\PT^\mathrm{UE}}$ & $\avg{
    \PT^\mathrm{ij}}$ & $\avg{\PT^\mathrm{ijl}}$ & $\rho(R)$ \\ \hline

  Track selection   & $<$0.2\% & $<$0.2\% & $<$0.2\% & $<$0.4\% & $<$1\%   \\
  Tracking performance & $<$0.3\% & $<$0.3\% & $<$0.4\% & $<$0.4\% &  $<$4\%  \\
  Model dependence  & $<$0.5\% & $<$0.4\% & $<$0.5\% & $<$0.5\% & $<$5\%   \\  \hline
  Statistical       & $<$0.1\% & $<$0.1\% & $<$0.2\% & $<$0.4\% & 2--8\%  \\ \hline
  Total             & 0.5--0.7\% & 0.5--0.6\% & 0.5--0.7\% & $<$0.9\% & 4--9\%  \\ \hline
\end{tabular}
\end{table*}

\begin{table*}
\topcaption{Summary of systematic and statistical uncertainties for various charged-jet related quantities.} \label{table:SumSysError2}
\centering
\begin{tabular}{lccccc}
\hline
              & ch. jet       & ch. jet rate &ch. jet rate & $\langle \PT^\text{ch. jet}\rangle$ \\
              &$\PT$ spectrum & ($\PT>5\GeVc$) & ($\PT>30$\GeVc) &           \\ \hline

  Track selection   & $<$1\%                               & $<$2\%   & $<$4\% & $<$0.1\%   \\

  Tracking performance & $<$3\%                               & 2\%    & $<$5\% &  $<$0.5\%  \\

  Model dependence  & $<$3\%                                &  2\%   & $<$6\% & $<$0.4\%   \\

  Unfolding         & 3\%                                &  $<$2\%  & $<$3\% &   $<$0.2\%\\   \hline

  Statistical       & 1--8\%                                                              & $<$1\% & $<$9\% & $<$0.4\%   \\
  		        & ($\PT^\text{ch. jet}<25$\GeVc)                      &        &        &            \\
                          & 10--40\%                                                          &        &        &            \\
                          & ($\PT^\text{ch. jet}>25$\GeVc)                      &        &        &            \\ \hline

  Total              &4--10\%                                                             & $<$5\%   & $<$12\% &  0.8\%  \\
                          &($\PT^\text{ch. jet}<25$\GeVc)                       &          &         &           \\
                         &10--40\%                                                           &          &         &           \\
                         &($\PT^\text{ch. jet}>25$ \GeVc)                       &          &         &           \\ \hline
\end{tabular}
\end{table*}

\section{Results}
\label{results}

\subsection{General properties of charged particles from jets and from the UE} \label{sec:GeneralProperties}

We start with discussing the general jet and UE properties in the five $N_\text{ch}$ bins defined.
Tables~\ref{global_properties},~\ref{global_properties2} list the average transverse momentum for the various
types of charged particles measured, as well as the predictions from \PYTHIA~8 tune 4C, \PYTHIA~8 MPI-off,
\PYTHIA~6 tune Z2*, and {\HERWIG}++ 2.5. For each multiplicity bin, we show the fully corrected results for the mean
transverse momenta of all charged particles $\avg{\PT^\text{ch. particle}}$,
UE charged particles $\avg{\PT^\mathrm{UE}}$, intrajet charged particles $\avg{
\PT^\mathrm{ij}}$, intrajet leading charged particles $\avg{\PT^\mathrm{ijl}}$, the mean transverse momentum
of charged-particle jets $\avg{\PT^\text{ch. jet}}$, and the average number of jets per event
$\avg{\frac{\#\text{jets}}{\text{event}}}$.

\begin{table*}
\centering
\topcaption{
Average transverse momenta for different types of charged particles (inclusive, underlying event, intrajet, leading intrajet).
The quantities are compared with the MC predictions. Uncertainties smaller than the last significant
digit are omitted.}
 \label{global_properties}
\begin{tabular}{cccccc}
\hline
   &  $\langle \PT^\text{ch. particle} \rangle $,\GeVc  &
   $\langle \PT^\mathrm{UE} \rangle$,\GeVc &
   $\langle \PT^\mathrm{ij} \rangle $,\GeVc  &
   $\langle \PT^\mathrm{ijl} \rangle $,\GeVc   \\ \hline\hline

\multicolumn{5}{c}{$10< N_\text{ch}\leq 30$} \\ \hline
    Data       &0.68 $\pm$ 0.01  & 0.65 $\pm$ 0.01 & 1.90 $\pm$ 0.02 & 3.65 $\pm$ 0.05 \\
    \PYTHIA 8 4C        & 0.67  &  0.64 & 1.83  &  3.48 $\pm$ 0.01 \\
    \PYTHIA 8 MPI-off & 0.72  &  0.66 & 1.93  &   3.73 \\
    \PYTHIA 6 Z2*       & 0.67  &   0.65 & 1.86 &   3.59 \\
    {\HERWIG}++ 2.5  & 0.68  &  0.65 &  1.81 &   3.41\\ \hline
\multicolumn{5}{c}{$30< N_\text{ch}\leq 50$}  \\ \hline
    Data       & 0.75 $\pm$ 0.01  & 0.71 $\pm$ 0.01 & 1.64 $\pm$ 0.02 & 3.37 $\pm$ 0.04 \\
    \PYTHIA 8 4C        & 0.77  &  0.72 & 1.62  & 3.25 $\pm$ 0.01 \\
    \PYTHIA 8 MPI-off & 1.06  &  0.75 & 1.99 & 4.28 $\pm$ 0.02 \\
    \PYTHIA 6 Z2*       & 0.74  &  0.70 &  1.62 & 3.33 \\
    {\HERWIG}++ 2.5  & 0.72  &  0.68 & 1.62 &  3.26\\ \hline
\multicolumn{5}{c}{$50< N_\text{ch}\leq 80$} \\ \hline
     Data       & 0.80 $\pm$ 0.01  &  0.74 $\pm$ 0.01 & 1.45 $\pm$ 0.01 & 3.15$\pm$ 0.03\\
    \PYTHIA 8 4C         & 0.84  &  0.76 & 1.49  & 3.10 \\
    \PYTHIA 8 MPI-off  & 1.47  &  0.80 &  2.22 & 5.17 $\pm$  0.09 \\
    \PYTHIA 6 Z2*        & 0.80  &  0.74 & 1.44 & 3.10\\
    {\HERWIG}++ 2.5   & 0.74  &  0.68 & 1.43 & 3.08 \\ \hline
\multicolumn{5}{c}{$80< N_\text{ch}\leq 110$} \\ \hline
    Data        & 0.85 $\pm$ 0.01  &  0.76 $\pm$ 0.01 & 1.32 $\pm$ 0.01 & 2.96 $\pm$ 0.03 \\
    \PYTHIA 8 4C         & 0.90  & 0.78  & 1.41 & 3.04 $\pm$ 0.01 \\
    \PYTHIA 6 Z2*        & 0.85  &  0.76 & 1.33 &  2.97 \\
    {\HERWIG}++ 2.5   & 0.74  &  0.66 & 1.28  & 2.94 \\ \hline
\multicolumn{5}{c}{$110< N_\text{ch}\leq 140$} \\ \hline
    Data       & 0.88 $\pm$ 0.01 &  0.77 $\pm$ 0.01 & 1.24 $\pm$ 0.01 & 2.86 $\pm$ 0.03\\
    \PYTHIA 8 4C        & 0.95  &  0.79 & 1.36 & 3.05 \\
    \PYTHIA 6 Z2*       & 0.90  &  0.77 & 1.29 & 3.05 $\pm$ 0.01\\
    {\HERWIG}++ 2.5 & 0.70   &  0.62 & 1.16 & 2.82 $\pm$ 0.01 \\ \hline
\end{tabular}
\end{table*}

\begin{table*}
\centering
\topcaption{
Average transverse momentum of charged-particle jets and charged-particle jet rates for two thresholds, $\PT>5\GeVc$ and $\PT>$~30\GeVc .
The quantities are compared with the MC predictions. Uncertainties smaller than the last significant
digit are omitted.}
 \label{global_properties2}
\begin{tabular}{cccc}
\hline
   &
   $\langle \PT^\text{ch. jet}\rangle$,\GeVc &
   $\langle \frac{\#\text{ch. jets}}{\text{event}} \rangle $  ($\PT^\text{ch. jet}>5\GeVc$)  &
   $\langle \frac{\#\text{ch. jets}}{\text{event}} \rangle $  ($\PT^\text{ch. jet}>30$\GeVc) \\ \hline\hline

\multicolumn{4}{c}{$10< N_\text{ch}\leq 30$} \\ \hline
    Data       & 6.85 $\pm$ 0.06 & 0.054 $\pm$ 0.004 &  (3.2$\pm$0.5)$ 10^{-5}$ \\
    \PYTHIA 8 4C         & 7.08 $\pm$ 0.01 & 0.075 &       (3.9$\pm$0.6)$ 10^{-5}$ \\
    \PYTHIA 8 MPI-off  & 7.96 $\pm$ 0.01&  0.152 &   (2.03$\pm$0.02)$ 10^{-4}$\\
    \PYTHIA 6 Z2*        & 7.01 $\pm$ 0.01 & 0.067 &    (2.7$\pm$0.3)$ 10^{-5}$ \\
    {\HERWIG}++ 2.5   & 6.92 $\pm$ 0.01 & 0.095 &    (3.8$\pm$0.5)$ 10^{-5}$ \\ \hline
\multicolumn{4}{c}{$30< N_\text{ch}\leq 50$} \\ \hline
    Data       & 7.04 $\pm$ 0.09 & 0.287 $\pm$ 0.014 &   (3.4$\pm$0.4)$ 10^{-4}$  \\
    \PYTHIA 8 4C         & 7.26 $\pm$ 0.01 & 0.386 &   (4.4$\pm$0.5)$ 10^{-4}$ \\
    \PYTHIA 8 MPI-off  &  10.8                     & 1.38 $\pm$ 0.02 &  (2.9$\pm$0.1)$ 10^{-2}$  \\
    \PYTHIA 6 Z2*        & 7.20 $\pm$ 0.01 & 0.304 &   (3.5$\pm$0.2)$ 10^{-4}$  \\
    {\HERWIG}++ 2.5   & 7.02 $\pm$ 0.01 & 0.375 &   (3.1$\pm$0.3)$ 10^{-4}$ \\ \hline
\multicolumn{4}{c}{$50< N_\text{ch}\leq 80$} \\ \hline
     Data       & 7.18  $\pm$ 0.09 & 0.84 $\pm$ 0.03 &  (1.5$\pm$0.1)$ 10^{-3}$ \\
    \PYTHIA 8 4C         & 7.41 $\pm$ 0.01 & 1.09 &       (1.8$\pm$0.1)$ 10^{-3}$ \\
    \PYTHIA 8 MPI-off  &  16.3 $\pm$ 0.4 & 3.1$\pm$ 0.3 &  (3.7$\pm$0.1)$ 10^{-1}$ \\
    \PYTHIA 6 Z2*        & 7.30 $\pm$ 0.01 & 0.87  &    (1.4$\pm$0.1)$ 10^{-3}$ \\
    {\HERWIG}++ 2.5   & 7.10 $\pm$ 0.01 & 0.88  &     (5.9$\pm$0.5)$ 10^{-4}$\\ \hline
\multicolumn{4}{c}{$80< N_\text{ch}\leq 110$} \\ \hline
    Data        & 7.46 $\pm$ 0.11 & 2.13 $\pm$ 0.09 &   (4.3$\pm$0.4)$ 10^{-3}$ \\
    \PYTHIA 8 4C         & 7.77 $\pm$ 0.02 & 2.54  &   (7.1$\pm$0.6)$ 10^{-3}$ \\
    \PYTHIA 6 Z2*        & 7.64 $\pm$ 0.01 & 2.12  &    (5.7$\pm$0.2)$ 10^{-3}$\\
    {\HERWIG}++ 2.5   & 7.25 $\pm$ 0.01 & 1.66  &    (1.2$\pm$0.1)$ 10^{-3}$\\ \hline
\multicolumn{4}{c}{$110< N_\text{ch}\leq 140$} \\ \hline
    Data       & 7.81 $\pm$ 0.10 & 3.68 $\pm$ 0.15 &   (1.0$\pm$0.1)$ 10^{-2}$ \\
    \PYTHIA 8 4C        & 8.31 $\pm$ 0.03 & 4.46  &  (2.5$\pm$0.1)$ 10^{-2}$ \\
    \PYTHIA 6 Z2*       & 8.15 $\pm$ 0.02 & 3.95  &   (2.1$\pm$0.1)$ 10^{-2}$\\
    {\HERWIG}++ 2.5  & 7.37 $\pm$ 0.01 & 2.41 &    (1.9$\pm$0.2)$ 10^{-3}$\\ \hline
\end{tabular}
\end{table*}

\begin{figure}[hbtp]
\begin{center}
\includegraphics[ width=\cmsFigWidth]{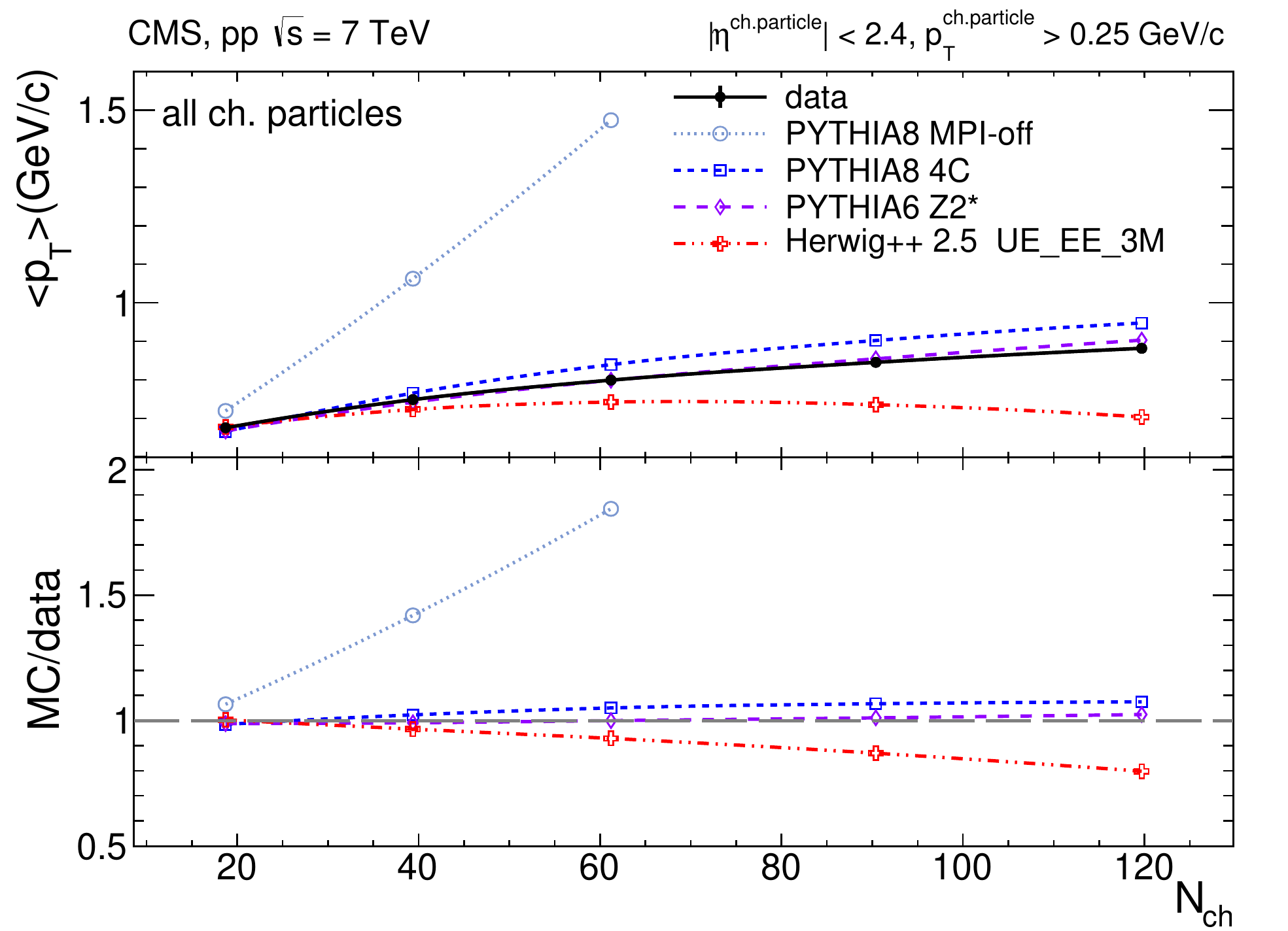}%
\caption{Mean transverse momentum of inclusive charged-particles with $\PT>0.25\GeVc$ versus
charged-particle multiplicity ($N_\text{ch}$ within $\abs{\eta}<$~2.4) measured in the data (solid line and marker)
compared to various MC predictions (non-solid curves and markers).
Systematic uncertainties are indicated by error bars which are, most of the time, smaller than the marker size.}%
\label{global_spectrum_vs_mult}%
\end{center}
\end{figure}

\begin{figure}[hbtp]
\begin{center}
\includegraphics[ width=\cmsFigWidth]{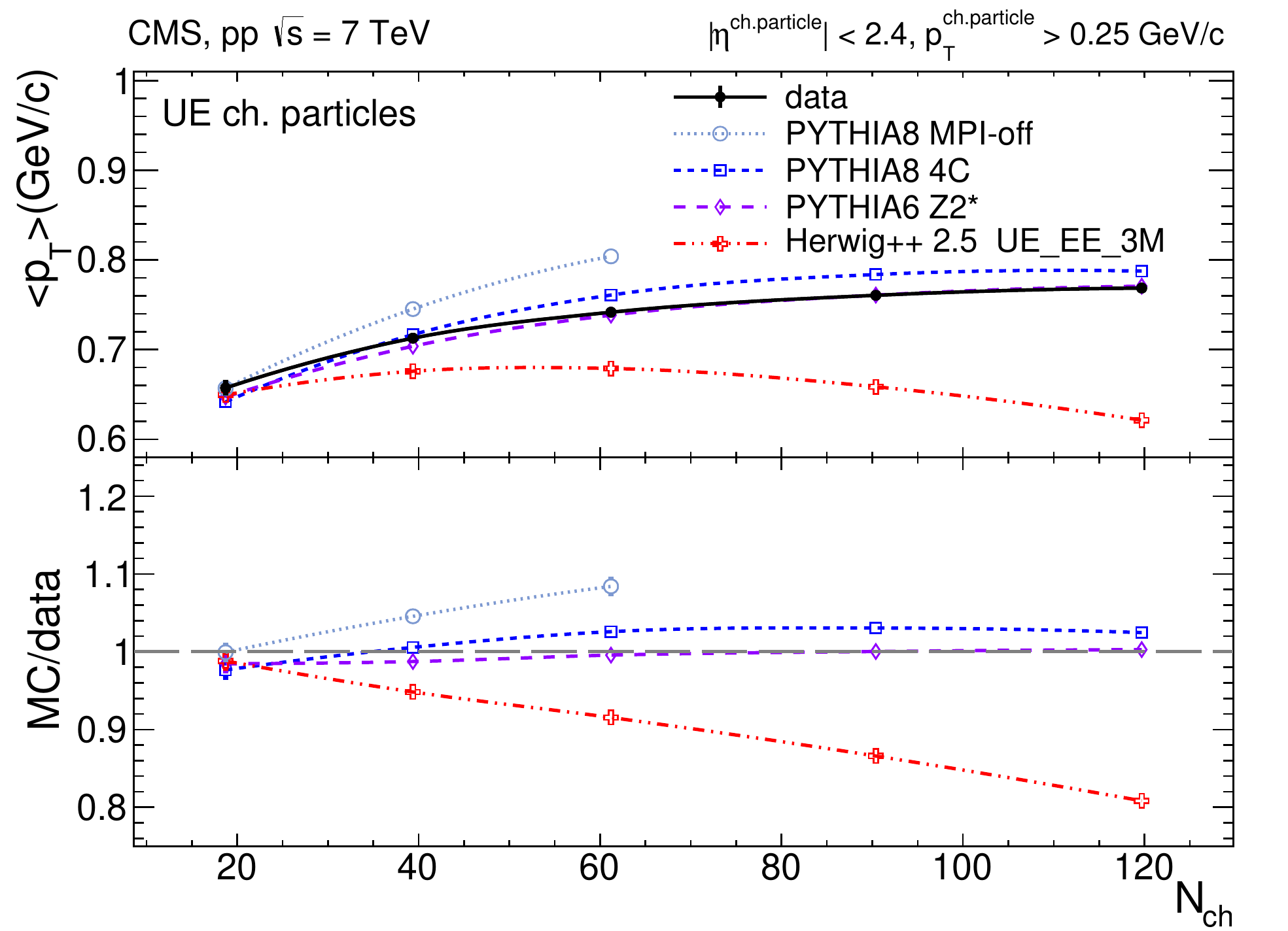}%
\caption{Mean transverse momentum of UE charged-particles with $\PT>0.25\GeVc$ versus
 charged-particle multiplicity ($N_\text{ch}$ within $\abs{\eta}<$~2.4) measured in the data (solid line and marker)
compared to various MC predictions (non-solid curves and markers).
Systematic uncertainties are indicated by error bars which are, most of the time, smaller than the marker size.}%
\label{soft_spectrum_vs_mult}%
\end{center}
\end{figure}

\begin{figure}[hbtp]
\begin{center}
\includegraphics[ width=\cmsFigWidth]{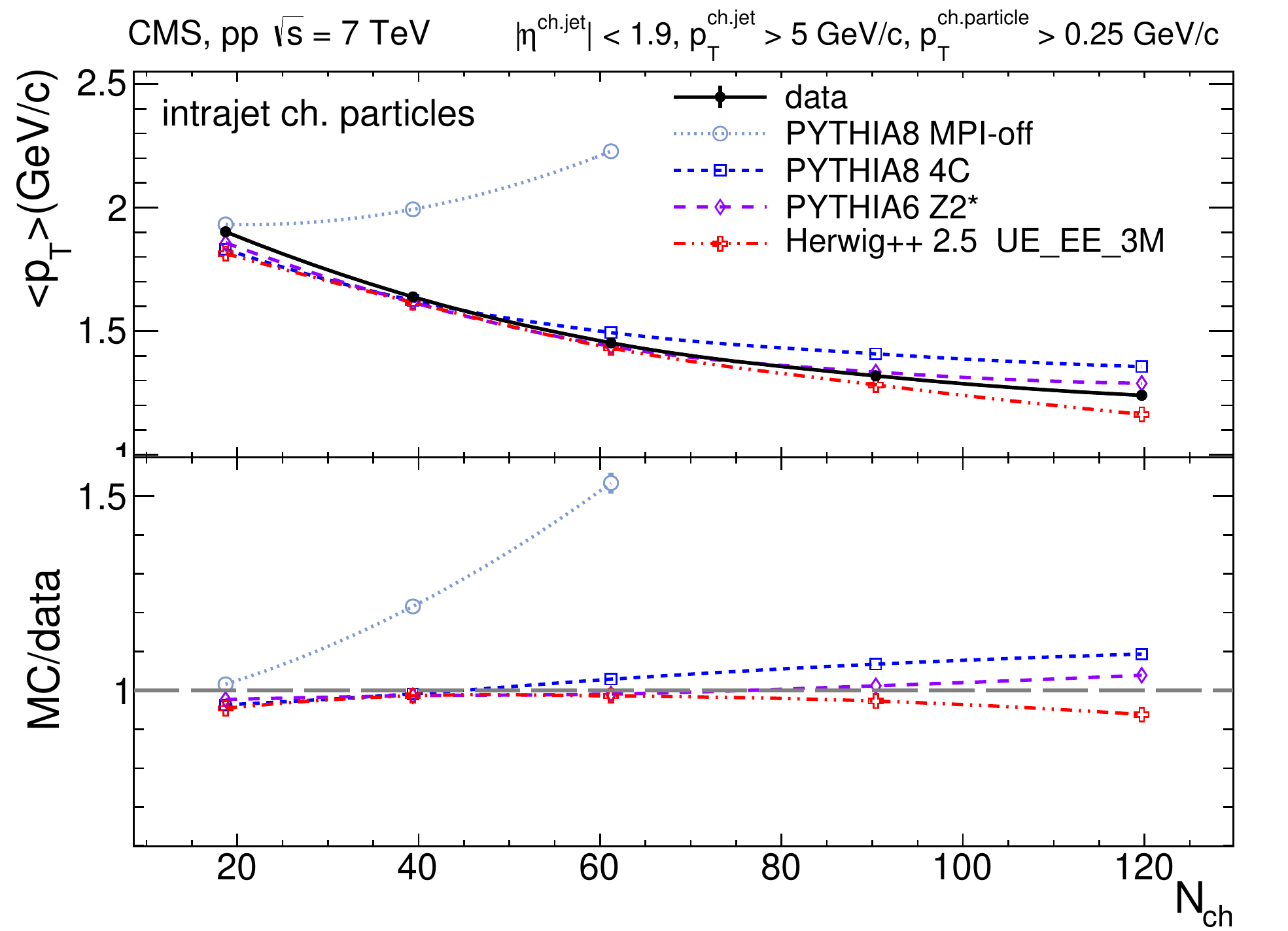}%
\caption{Mean transverse momentum of intrajet charged-particles with $\PT>0.25\GeVc$ versus
 charged-particle multiplicity ($N_\text{ch}$ within $\abs{\eta}<2.4$) measured in the data (solid line and marker)
compared to various MC predictions (non-solid curves and markers).
Systematic uncertainties are indicated by error bars which are, most of the time, smaller than the marker size.}%
\label{intrajet_spectrum_vs_mult_0.4}%
\end{center}
\end{figure}

The mean transverse momenta of all charged particles, UE charged-particles, and intrajet charged-particles,
are plotted as a function of $N_\text{ch}$ in Figs. \ref{global_spectrum_vs_mult}--\ref{intrajet_spectrum_vs_mult_0.4}.
From Figs. \ref{global_spectrum_vs_mult} and \ref{soft_spectrum_vs_mult}, we see that mean transverse momentum
of inclusive and UE charged-particles increases with $N_\text{ch}$. Such a behavior is expected as the
higher multiplicity events have an increased fraction of (semi)hard scatterings contributing to final hadron production.
The (logarithmic-like) $N_\text{ch}$-dependence of the average transverse momentum of inclusive and UE
charged-particles is well described by both \PYTHIA~6 tune Z2* and \PYTHIA~8 tune 4C (especially by the
former), and is less well described by {\HERWIG}++  2.5, which does not predict a monotonically rising dependence
but a ``turn down'' beyond $N_\text{ch}\approx$~60.
On the other hand, \PYTHIA~8 without MPI fails to describe the data altogether, predicting much harder
charged-particle spectra for increasing final multiplicity. This follows from the fact that  \PYTHIA~8
without MPI can only produce high-multiplicity events through very hard jets with large intrajet multiplicity,
instead of producing a larger number of semi-hard jets in the event.

\begin{figure}[hbtp]
\begin{center}
\includegraphics[ width=\cmsFigWidth]{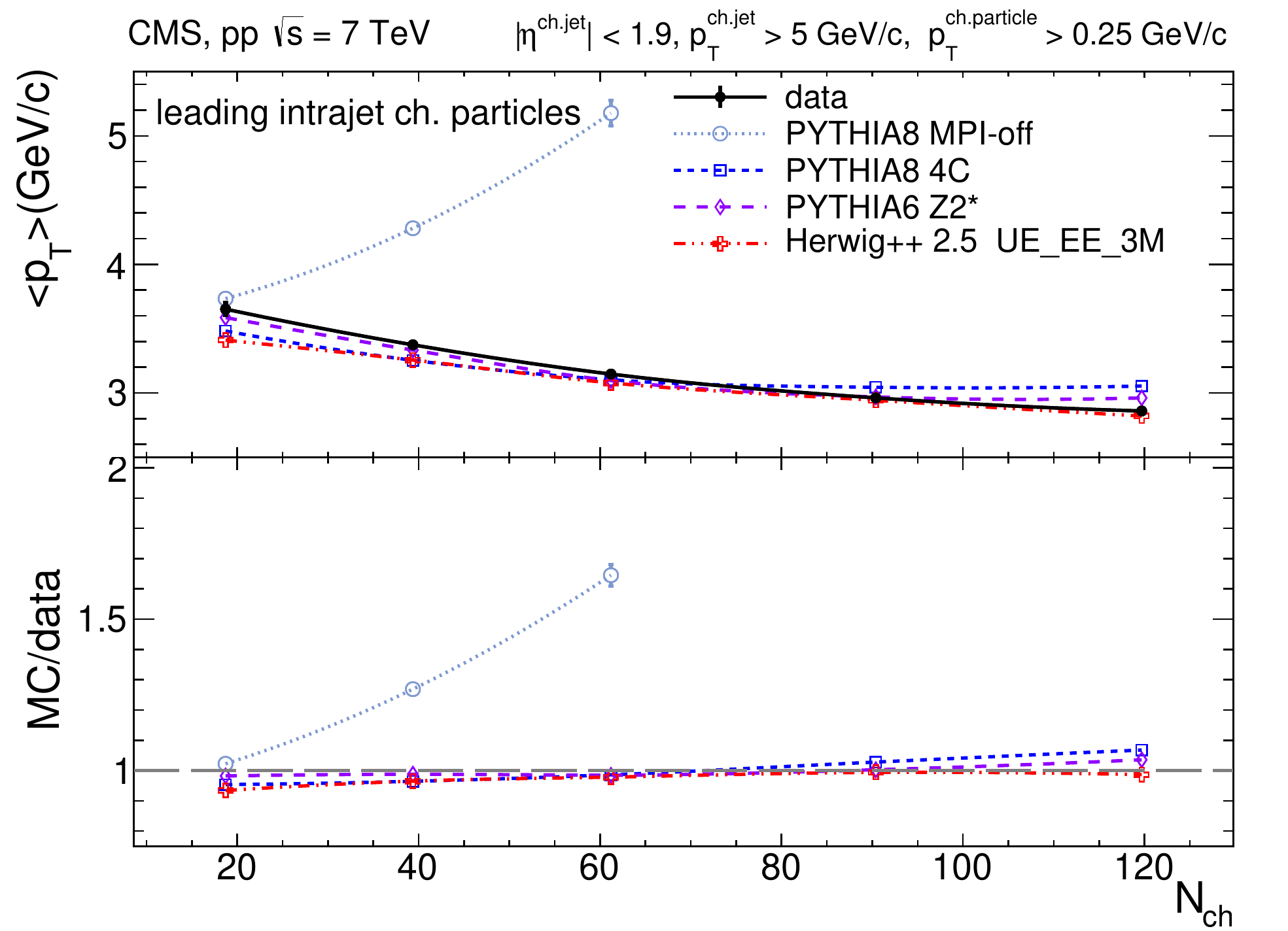}%
\caption{Mean transverse momentum of leading intrajet charged-particles with $\PT>0.25\GeVc$ versus
 charged-particle multiplicity ($N_\text{ch}$ within $\abs{\eta}<2.4$) measured in the data (solid line and marker)
compared to various MC predictions (non-solid curves and markers).
Systematic uncertainties are indicated by error bars which are, most of the time, smaller than the marker size.}%
\label{intrajet_leader_spectrum_vs_mult}%
\end{center}
\end{figure}

From Figs.~\ref{intrajet_spectrum_vs_mult_0.4}--\ref{intrajet_leader_spectrum_vs_mult} it is clear that the
$N_\text{ch}$-dependence of the average $\PT$ of intrajet constituents and leading charged-particle of the jets
shows the opposite behavior compared to that from the global and underlying events
(Figs. \ref{global_spectrum_vs_mult}--\ref{soft_spectrum_vs_mult}) and decreases logarithmically with
increasing multiplicities.
Events with increasing multiplicities are naturally ``biased'' towards final-states resulting
mostly from (mini)jets which fragment into a (increasingly) large number of hadrons. Since the produced hadrons
share the energy of the parent parton, a larger amount of them results in overall softer intrajet- and
leading-hadron $\PT$ spectra. Part of the decrease of the intrajet mean $\PT$ with multiplicity could be also due to
extra soft UE contribution falling within the jet cones, which increases from about 5\% for $N_\text{ch}\approx20$,
to about 20\% for $N_\text{ch}\approx120$, according to \PYTHIA~6 tune Z2*.
In terms of data-MC comparisons, we see that \PYTHIA~6 tune Z2* and {\HERWIG}++ 2.5 describe relatively well
the $N_\text{ch}$-dependence of the intrajet and leading-particle average $\PT$, whereas \PYTHIA~8 tune 4C
produces harder mean charged-particle spectra at high multiplicities.
The \PYTHIA~8 predictions
without MPI increase dramatically with $N_\text{ch}$, and fail to describe the data.
This can be explained by the fact that \PYTHIA  MPI-off enriches the
increasing multiplicity range with events with hard partons only, whereas the other MC models
include additional semi-hard parton interactions that soften the final hadron $\PT$ spectra.

\subsection{Charged-particle jet properties}

In the previous section, the jet substructure was investigated via the averaged properties of intrajet and leading
particles. Now we turn to the description of the multiplicity-dependent properties of the jets themselves.
In general, properties of inclusive jet production, when integrated over all multiplicities,
 are  dominated by events with moderately low multiplicities, and are described quite
well by QCD MC models~\cite{CMSevent, kodo,ATL2, ATL3}.
Here, we concentrate on the $N_\text{ch}$-dependence of a subset of jet properties, such as the number of jets per event,
the mean transverse momenta of jets, differential jet $\PT$ spectra, and jet widths.

Our study is complementary to others based
on global event shapes, \eg from the ALICE experiment~\cite{AL}, which observed an increasing
event transverse sphericity as a function of multiplicity in contradiction with the MC predictions.
However, the corresponding multiplicities are much lower in the ALICE study than
in this analysis because of their smaller rapidity coverage ($\abs{\eta} <0.8$).
Similar observations have been also recently seen by ATLAS~\cite{ATL}, even though earlier CMS and ATLAS
results show no serious disagreement with MC event generators~\cite{ATL2,CMSevent} as the events were not
sorted according to their multiplicity.
We show here that the higher sphericity of high-multiplicity events, relative to the \PYTHIA predictions, is
due to an apparent reduction and softening of the jet yields at high-$N_\text{ch}$.

\subsubsection{Charged-particle jet production rates}

\begin{figure}[hbtp]
\begin{center}
\includegraphics[ width=\cmsFigWidth]{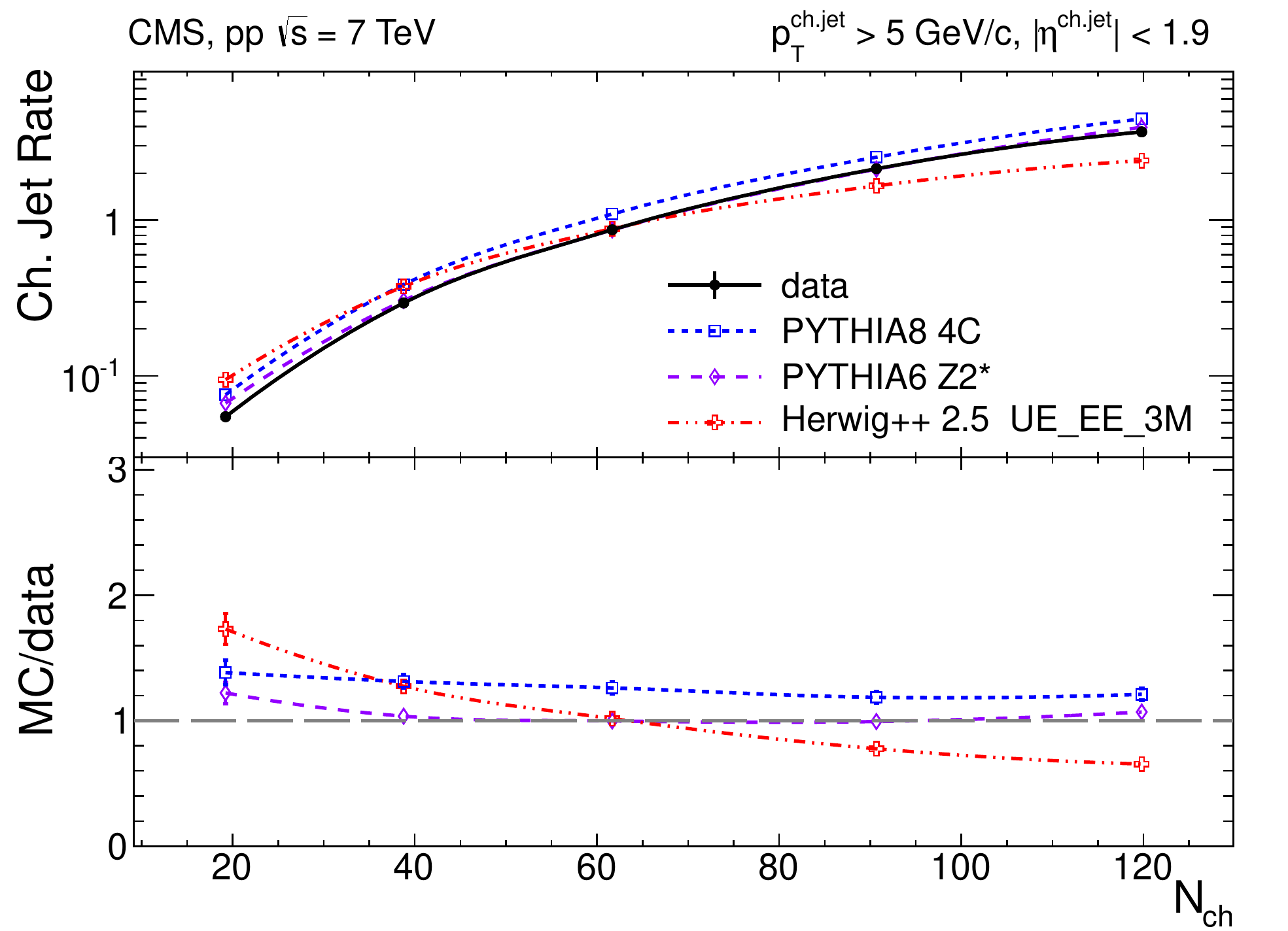}%
\caption{Number of charged-particle jets per event for $\PT^\text{ch. jet}>5\GeVc$ and jet axes lying within
$\abs{\eta}<1.9$ versus charged-particle multiplicity ($N_\text{ch}$ within $\abs{\eta}<2.4$) measured in the data
(solid line and marker) compared to various MC predictions (non-solid curves and markers).
Systematic uncertainties are indicated by error bars which are, most of the time, smaller than the marker size.}%
\label{JetRate_5GeV}%
\end{center}
\end{figure}

\begin{figure}[hbtp]
\begin{center}
\includegraphics[ width=\cmsFigWidth]{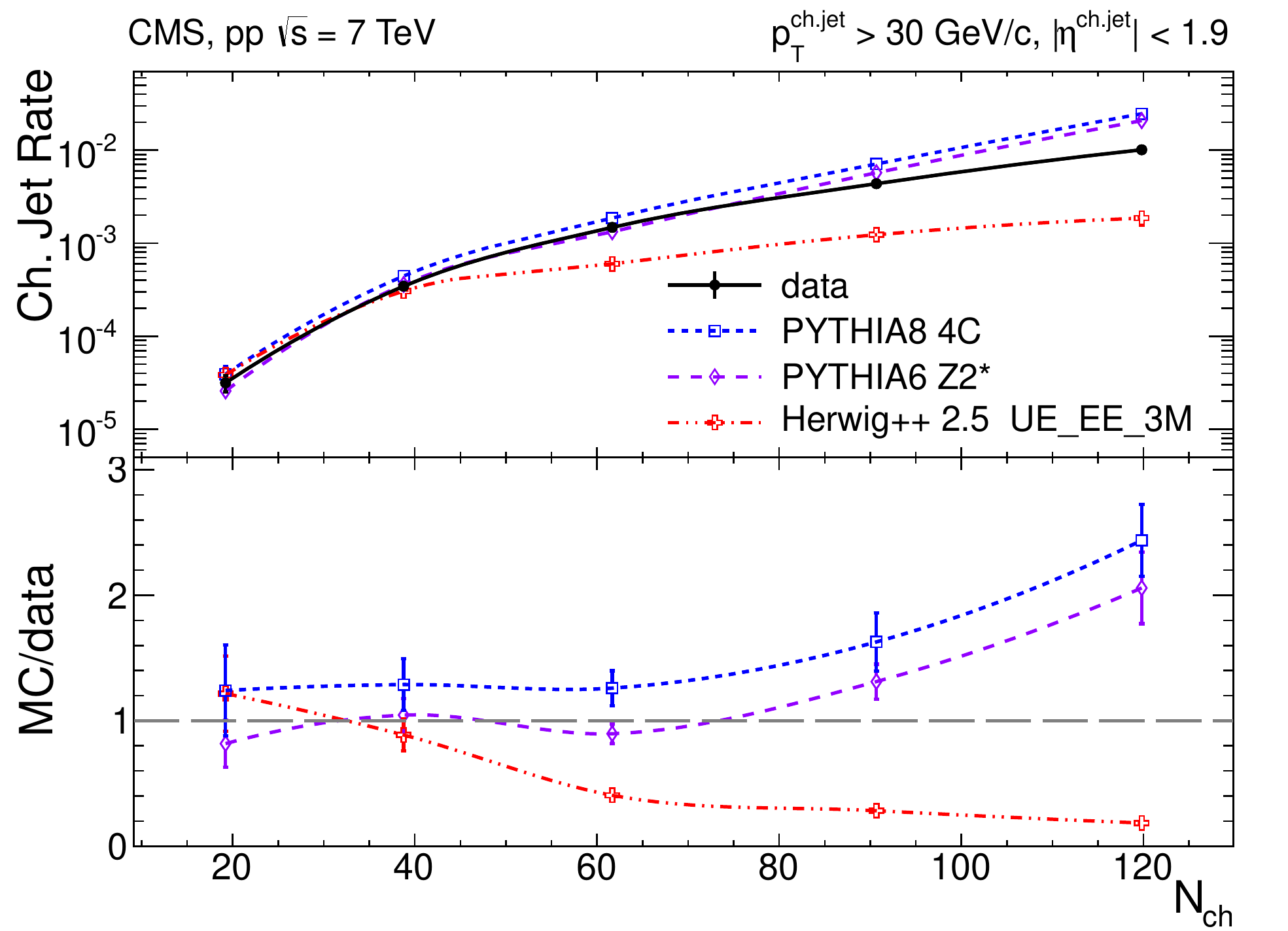}%
\caption{Number of charged-particle jets per event for $\PT^\text{ch. jet}>30\GeVc$ and jet axes lying within
$\abs{\eta}<1.9$ versus charged-particle multiplicity ($N_\text{ch}$ within $\abs{\eta}<2.4$) measured in the data
(solid line and marker) compared to various MC predictions (non-solid curves and markers).
Error bars denote the total uncertainties.}%
\label{JetRate_30GeV}%
\end{center}
\end{figure}

The $N_\text{ch}$-dependence of the number of jets per event, with jet transverse momentum
 $\PT^\text{ch. jet}>5$\GeVc and $\PT^\text{ch. jet}>30$\GeVc, is shown in
Figs.~\ref{JetRate_5GeV} and \ref{JetRate_30GeV}, respectively.

For the small cutoff of 5\GeVc the data show an increase from an average of 0.05~jets/event to about 4~jets/event going
from the lowest to the highest charged-particle multiplicities. Such results, which confirm the importance of
multiple (mini)jet production to explain the high-$N_\text{ch}$ events, are very well described by
\PYTHIA~6 tune Z2*, while predictions of \PYTHIA~8 tune 4C overestimate the rates at all $N_\text{ch}$ and \HERWIG++  2.5 underestimates them for increasing $N_\text{ch}$. For the higher 30\GeVc
cutoff, a large disagreement with the data is found in the higher-multiplicity bins
(Fig. \ref{JetRate_30GeV}), where both versions of \PYTHIA  predict a
factor of two more jets per event than seen in the data. On the contrary, {\HERWIG}++ 2.5  predicts a factor of 5
fewer jets per event than experimentally measured. The prediction of \PYTHIA~8 without MPI contributions
is completely off-scale by factors of 3.5--6 above the data and is not shown in the plots.

\begin{figure}[hbtp]
\begin{center}
\includegraphics[ width=\cmsFigWidth]{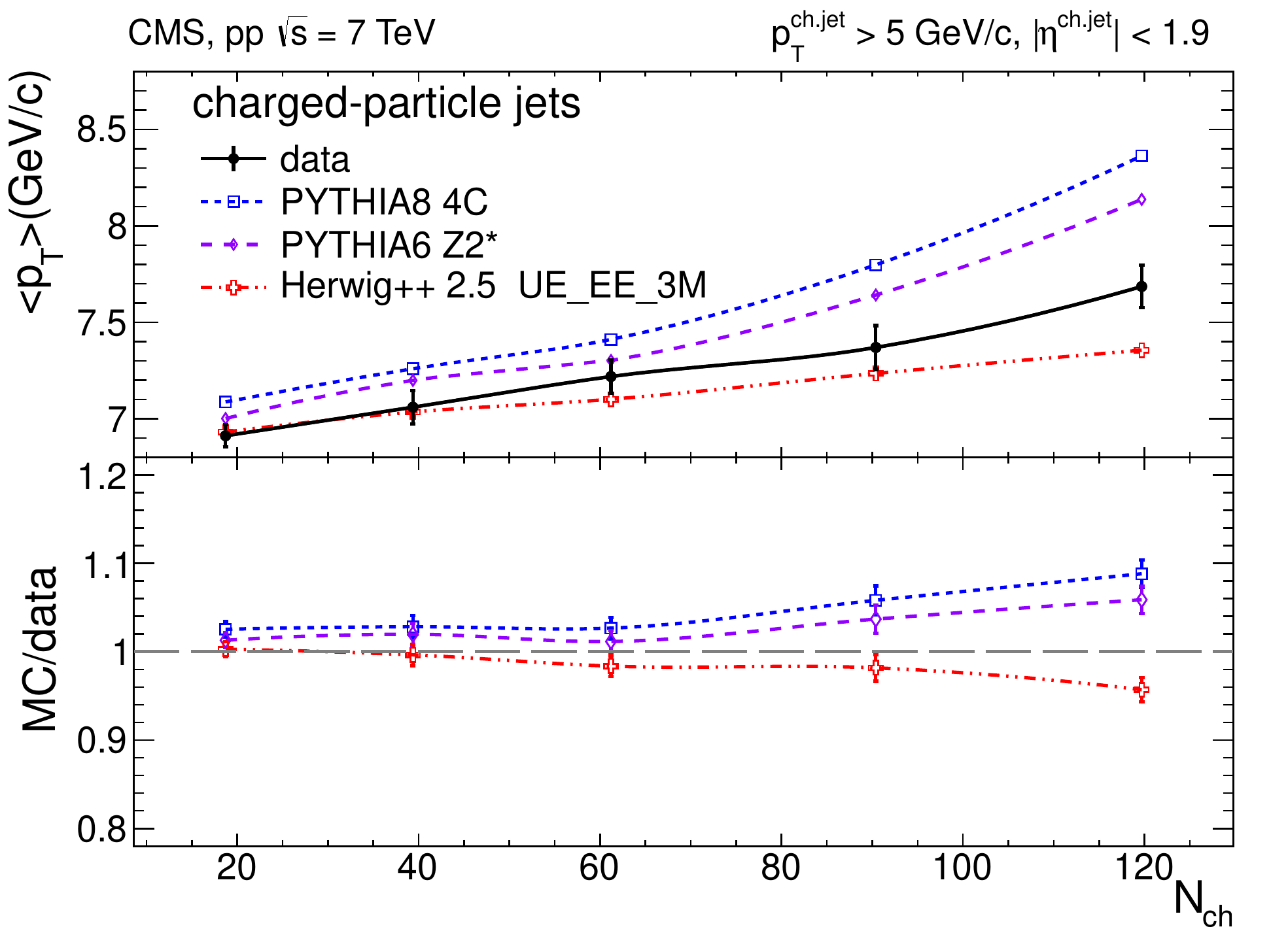}%
\caption{Mean transverse momentum of charged-particle jets with $\PT^\text{ch. jet}>5\GeVc$ and jet axes within
 $\abs{\eta}<1.9$) versus charged-particle multiplicity ($N_\text{ch}$ within $\abs{\eta}<2.4$) measured in the data
 (solid line and marker) compared to various MC predictions (non-solid curves and markers). Error bars denote
 the total uncertainties.}%
\label{jetptvsm}%
\end{center}
\end{figure}

The analysis of the $N_\text{ch}$-dependence of the mean transverse momentum of charged-particle jets
$\langle \PT^\text{ch. jet} \rangle$ is shown in Fig.~\ref{jetptvsm}. The average $\langle \PT^\text{ch. jet}
\rangle$ rises slowly with $N_\text{ch}$ from about 7.0 to 7.7\GeVc, indicating a rising contribution
from harder scatterings for increasingly ``central'' pp events.
The predictions of \PYTHIA~8 tune 4C, \PYTHIA~6 tune Z2*, and {\HERWIG}++ 2.5 are in good agreement with the
data at low and intermediate multiplicities. However, the \PYTHIA models display an increasingly higher value
of $\langle \PT^\text{ch. jet} \rangle$, i.e. a harder jet contribution, up to 8.4\GeVc in the
highest-multiplicity events.

\subsubsection{Charged-particle jet spectra}

\begin{figure}[hbtp]
\begin{center}
\includegraphics[ width=\cmsFigWidth]{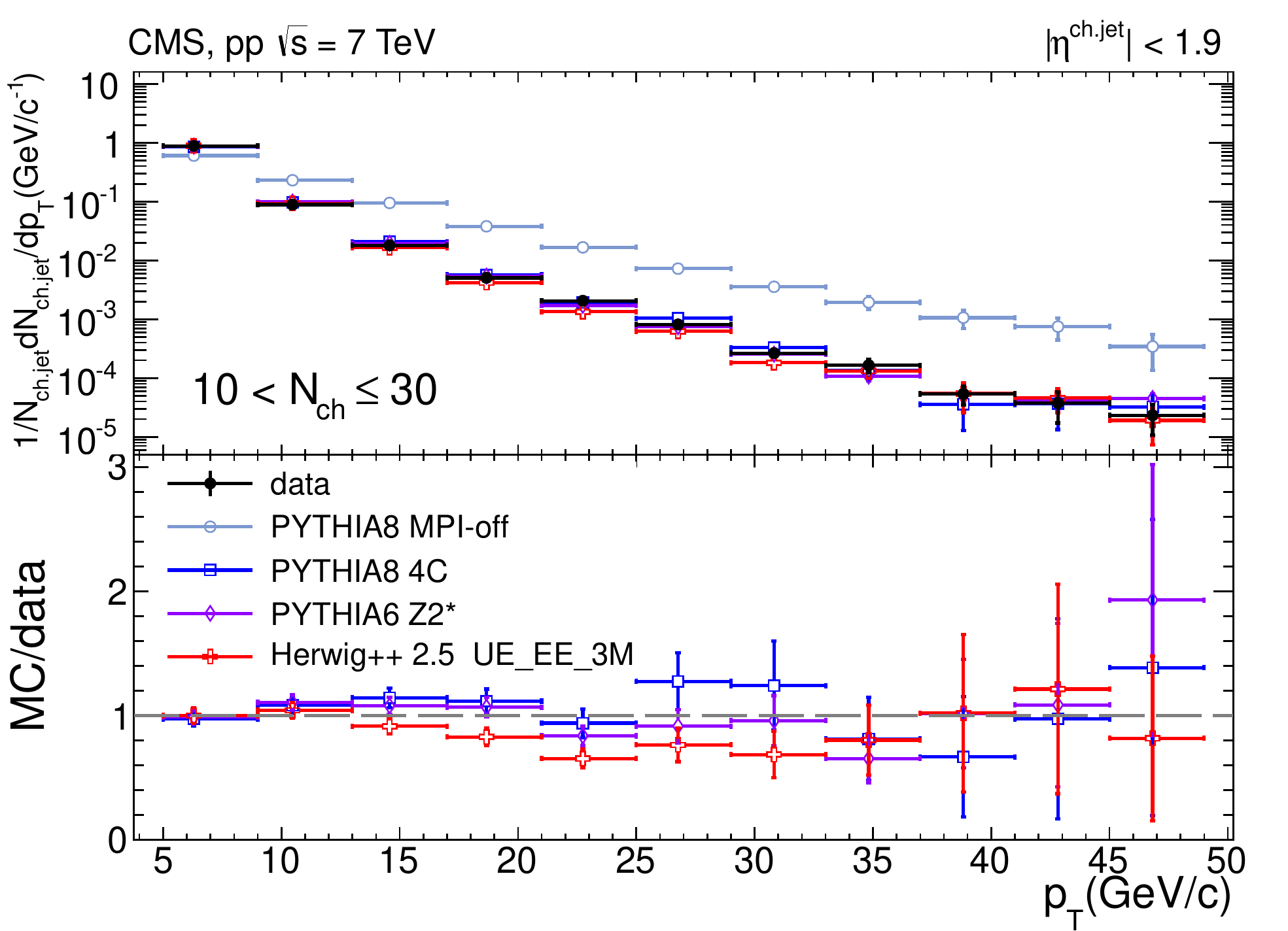}%
\caption{Inclusive charged-particle jet $\PT$ spectrum for events with $10<N_\text{ch}(\abs{\eta}<2.4)\leq 30$
 measured in the data (solid dots) compared to various MC predictions (empty markers).
Error bars denote the total uncertainties.}%
\label{jetpt1030}%
\end{center}
\end{figure}

\begin{figure}[hbtp]
\begin{center}
\includegraphics[ width=\cmsFigWidth]{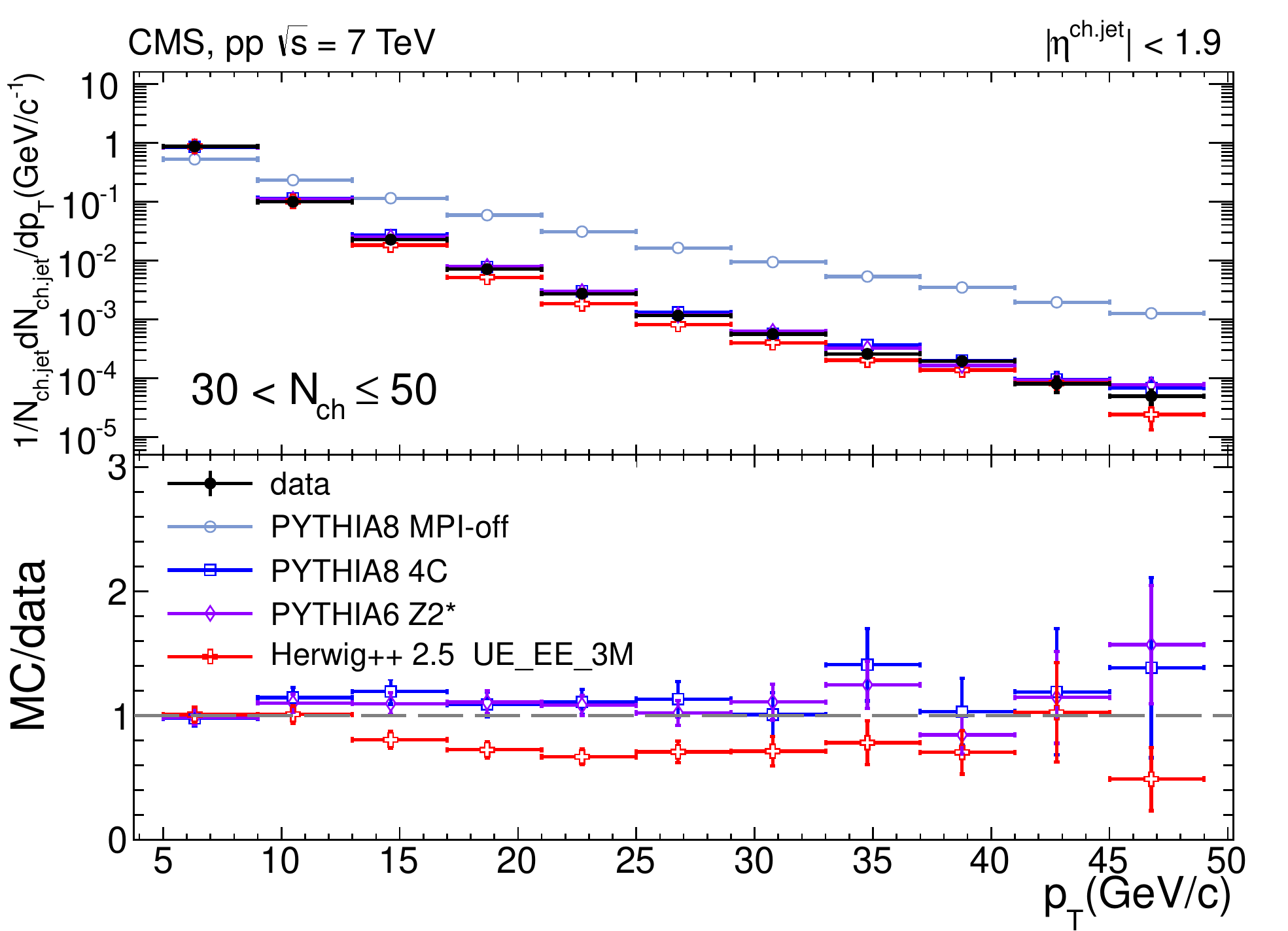}%
\caption{Inclusive charged-particle jet $\PT$ spectrum for events with $30<N_\text{ch}(\abs{\eta}<2.4)\leq 50$
 measured in the data (solid dots) compared to various MC predictions (empty markers).
Error bars denote the total uncertainties.}%
\label{jetpt3050}%
\end{center}
\end{figure}

\begin{figure}[hbtp]
\begin{center}
\includegraphics[ width=\cmsFigWidth]{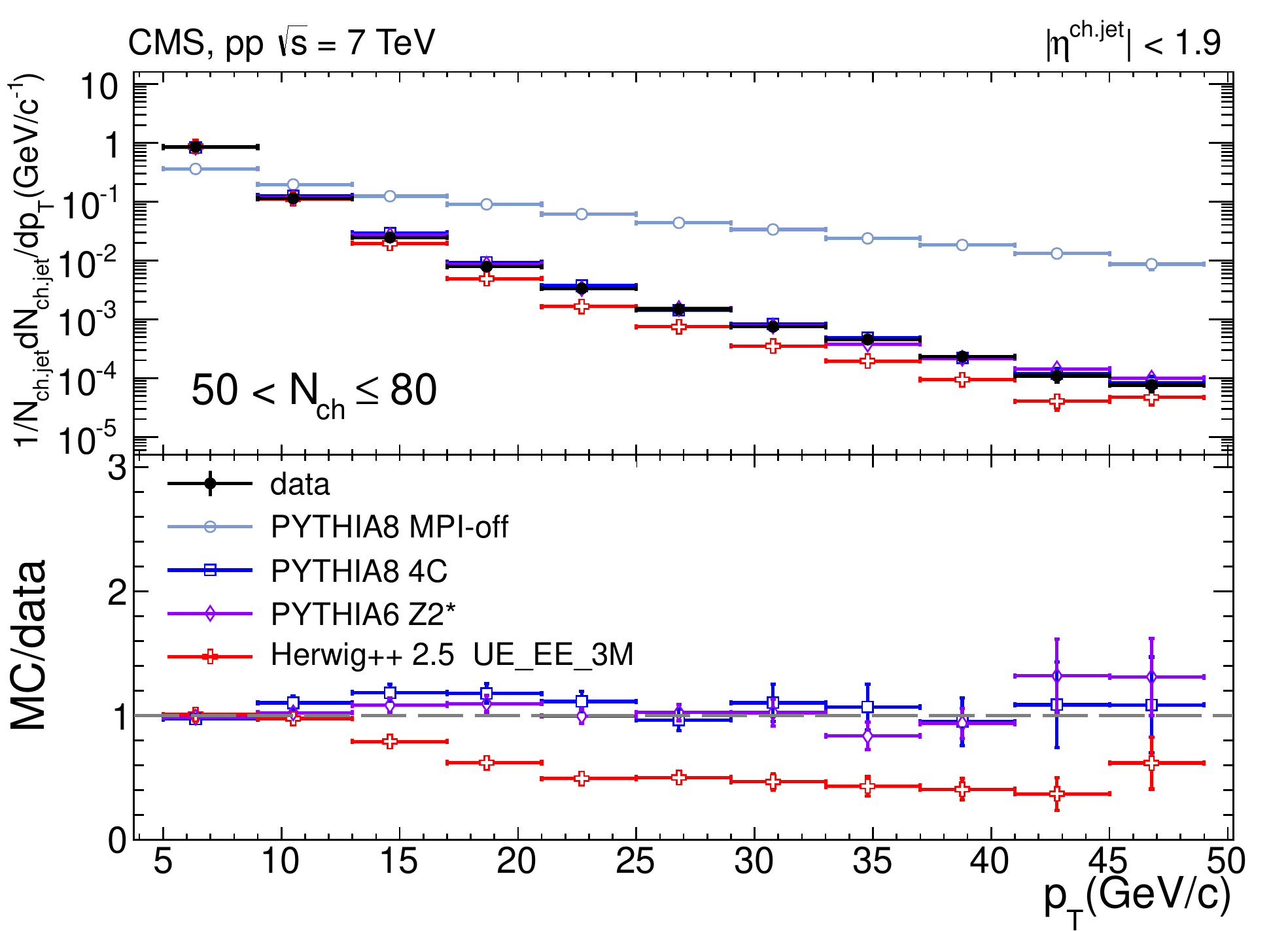}%
\caption{Inclusive charged-particle jet $\PT$ spectrum for events with $50<N_\text{ch}(\abs{\eta}<2.4) \leq 80$
 measured in the data (solid dots) compared to various MC predictions (empty markers).
Error bars denote the total uncertainties.}%
\label{jetpt5070}%
\end{center}
\end{figure}

\begin{figure}[hbtp]
\begin{center}
\includegraphics[ width=\cmsFigWidth]{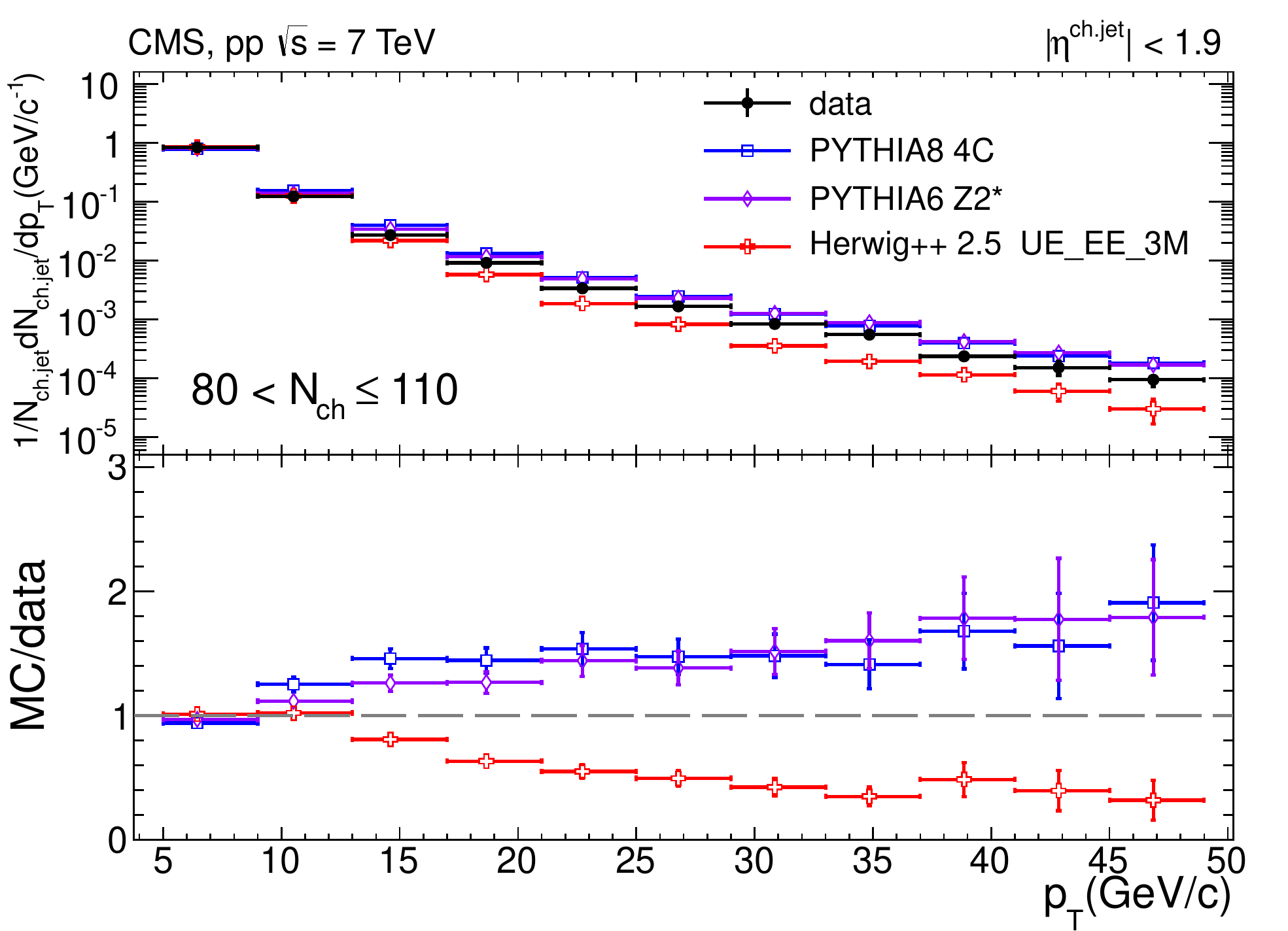}%
\caption{Inclusive charged-particle jet $\PT$ spectrum for events with $80<N_\text{ch}(\abs{\eta}<2.4) \leq 110$
 measured in the data (solid dots) compared to various MC predictions (empty markers).
Error bars denote the total uncertainties.}%
\label{jetpt70100}%
\end{center}
\end{figure}

\begin{figure}[hbtp]
\begin{center}
\includegraphics[ width=\cmsFigWidth]{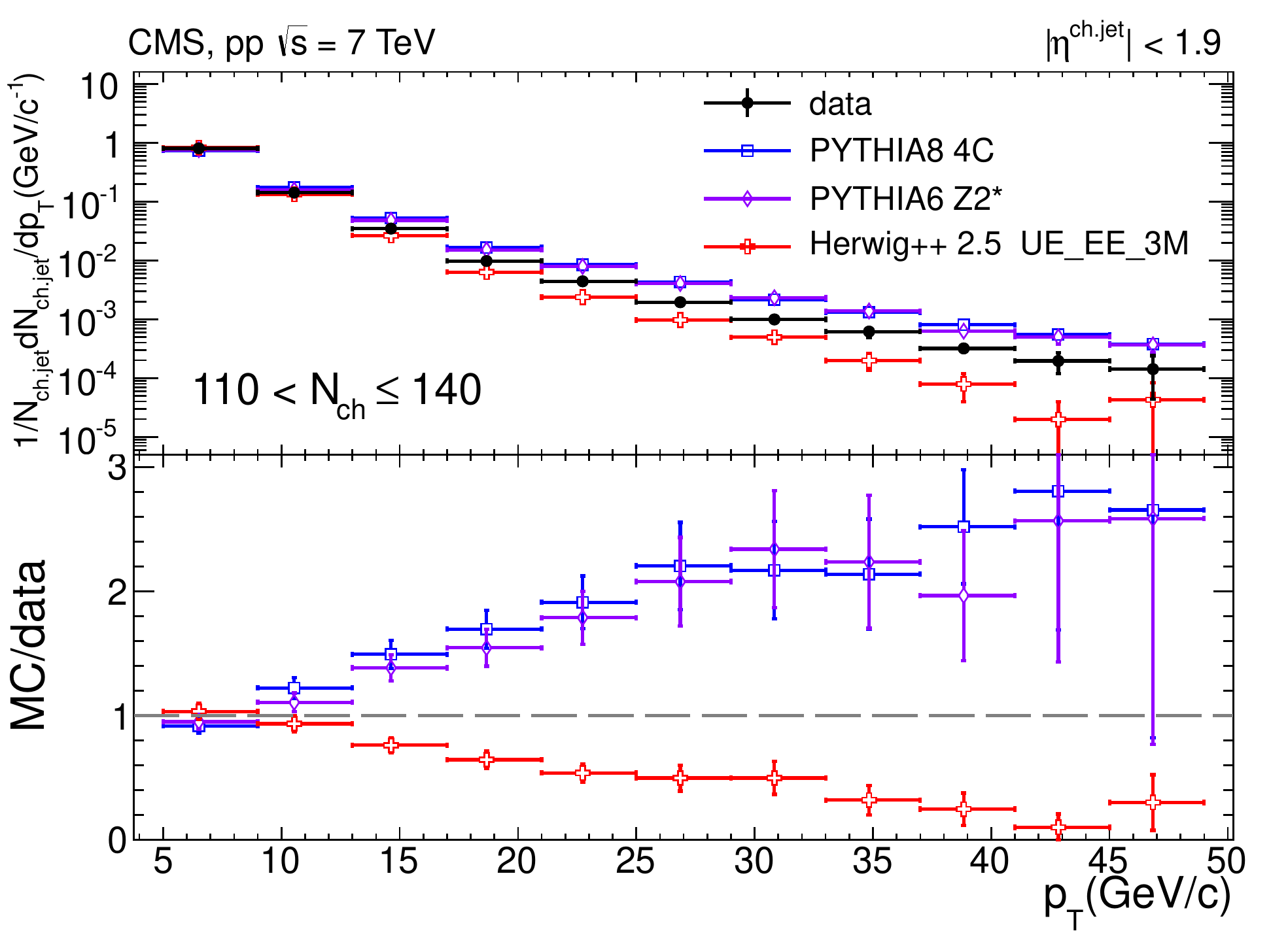}%
\caption{Inclusive charged-particle jet $\PT$ spectrum for events with $110 < N_\text{ch}(\abs{\eta}<2.4) \leq 140$
 measured in the data (solid dots) compared to various MC predictions (empty markers).
Error bars denote the total uncertainties.}%
\label{jetpt100130}%
\end{center}
\end{figure}

A more detailed picture of the properties of jet spectra both in data and MC simulations is provided by
directly comparing the $\PT$-differential distributions in each of the five multiplicity bins shown in
Figs.~\ref{jetpt1030}--\ref{jetpt100130}. In the first three $N_\text{ch}$ bins the measured jet $\PT$
spectra are reasonably well reproduced by the MC predictions. However, in the two highest-multiplicity bins,
$80< N_\text{ch} \leq 110$ (Fig.~\ref{jetpt70100}) and $110 < N_\text{ch} \leq 140$ (Fig.~\ref{jetpt100130}), we
observe much softer jet spectra for transverse momenta $\PT> 20$\GeVc, where data are lower by a
factor of $\sim$2 with respect to \PYTHIA predictions. At the same time,  \HERWIG++ 2.5  shows the opposite trend, and
predicts softer charged-particle jets than measured in data in all multiplicity bins. The relative ``softening'' of the
measured jet spectra compared to \PYTHIA at high-$N_\text{ch}$, explains also the higher sphericity of
high-multiplicity events observed in Ref.~\cite{AL}.

\subsubsection{Charged-particle jet widths}\label{sec:JetWidth}

\begin{figure}[hbtp]
\begin{center}
\includegraphics[width=\cmsFigWidth]{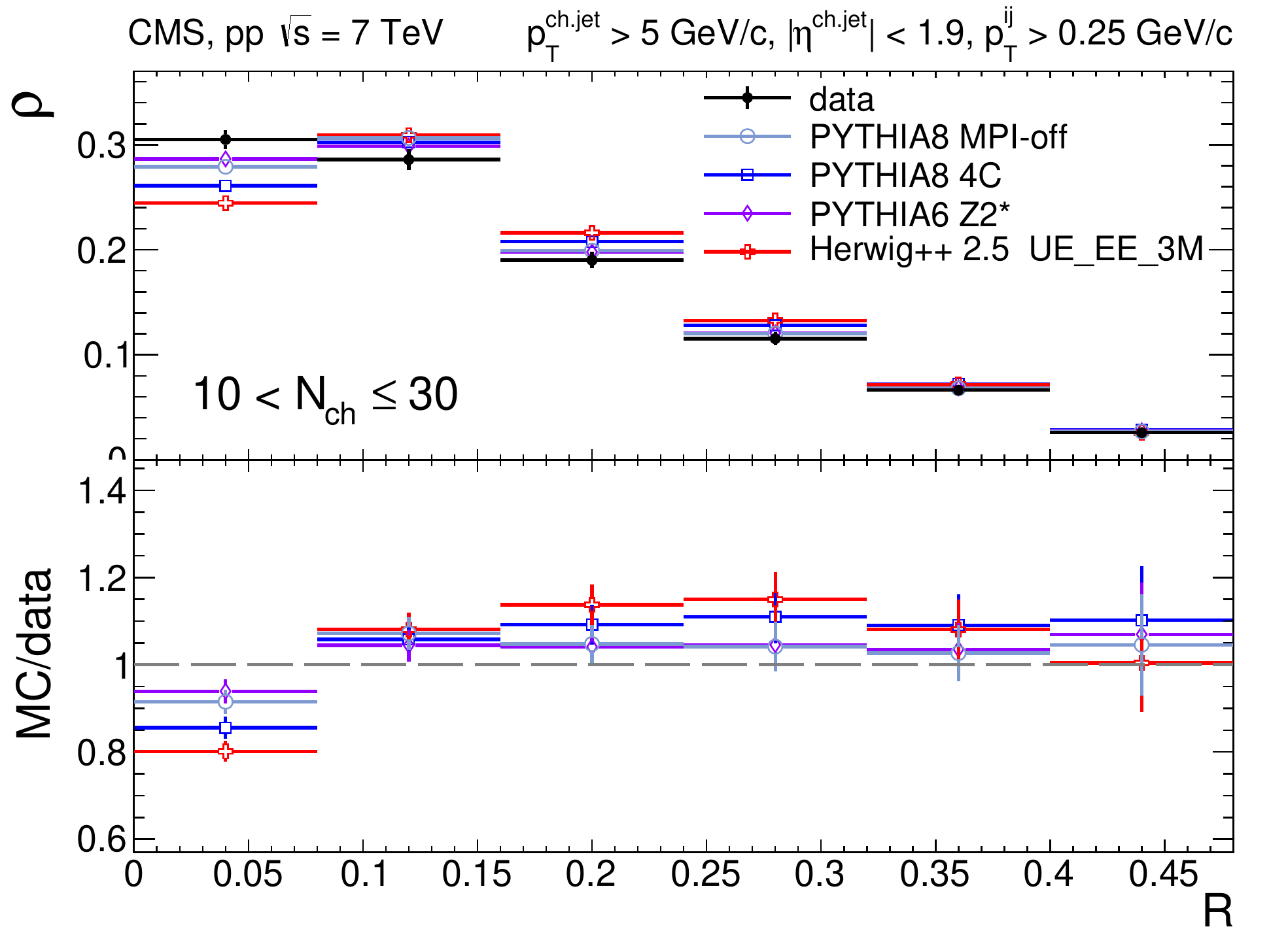}%
\caption{Normalized charged-particle jet $\PT$ density $\rho$ in ring zones as a function of distance to the jet axis $R$
  for events with $10< N_\text{ch}(\abs{\eta}<2.4) \leq 30$ measured in the data (solid dots) compared to
  various MC predictions (empty markers).
 Error bars denote the total uncertainties.}%
\label{jetstruct1030}%
\end{center}
\end{figure}

\begin{figure}[hbtp]
\begin{center}
\includegraphics[width=\cmsFigWidth]{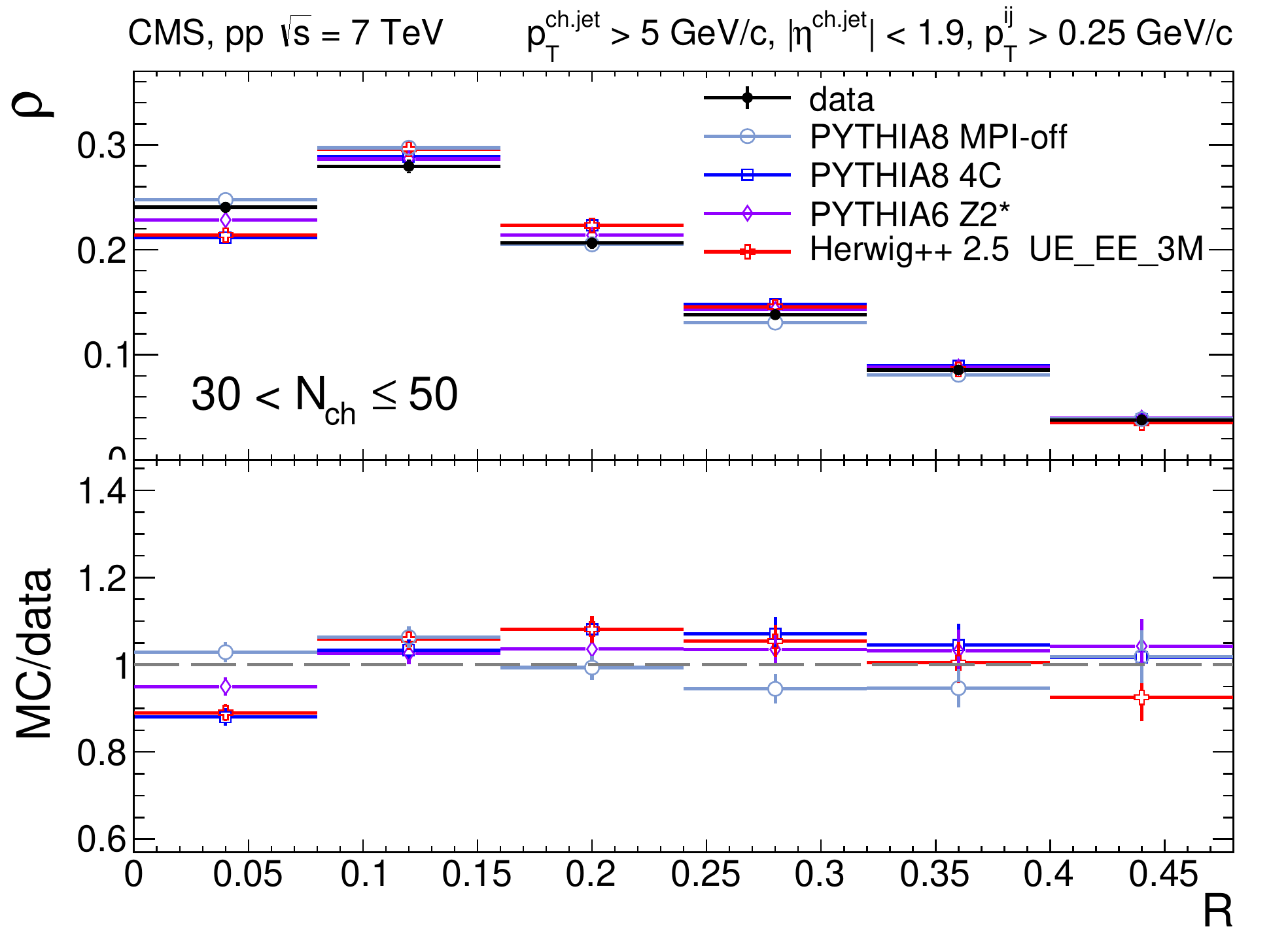}%
\caption{Normalized charged-particle jet $\PT$ density $\rho$ in ring zones as a function of distance to the jet axis $R$
  for events with $30< N_\text{ch}(\abs{\eta}<2.4) \leq 50$ measured in the data (solid dots) compared to
  various MC predictions (empty markers). Error bars denote the total uncertainties.}%
\label{jetstruct3050}%
\end{center}
\end{figure}

\begin{figure}[hbtp]
\begin{center}
\includegraphics[width=\cmsFigWidth]{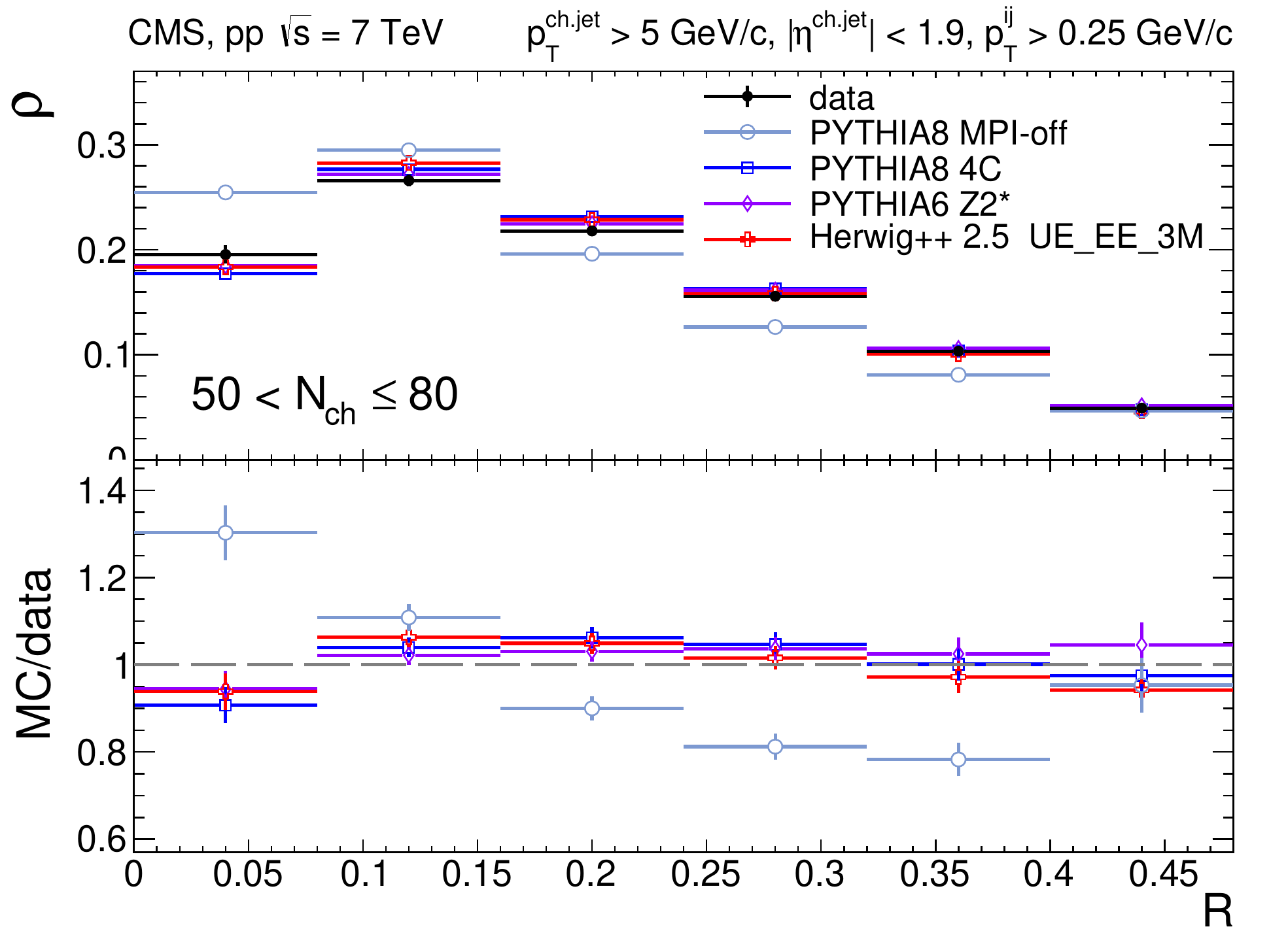}%
\caption{Normalized charged-particle jet $\PT$ density $\rho$ in ring zones as a function of distance to the jet axis $R$
  for events with $50< N_\text{ch}(\abs{\eta}<2.4) \leq 80$ measured in the data (solid dots) compared to
  various MC predictions (empty markers). Error bars denote the total uncertainties.}%
\label{jetstruct5070}%
\end{center}
\end{figure}

\begin{figure}[hbtp]
\begin{center}
\includegraphics[width=\cmsFigWidth]{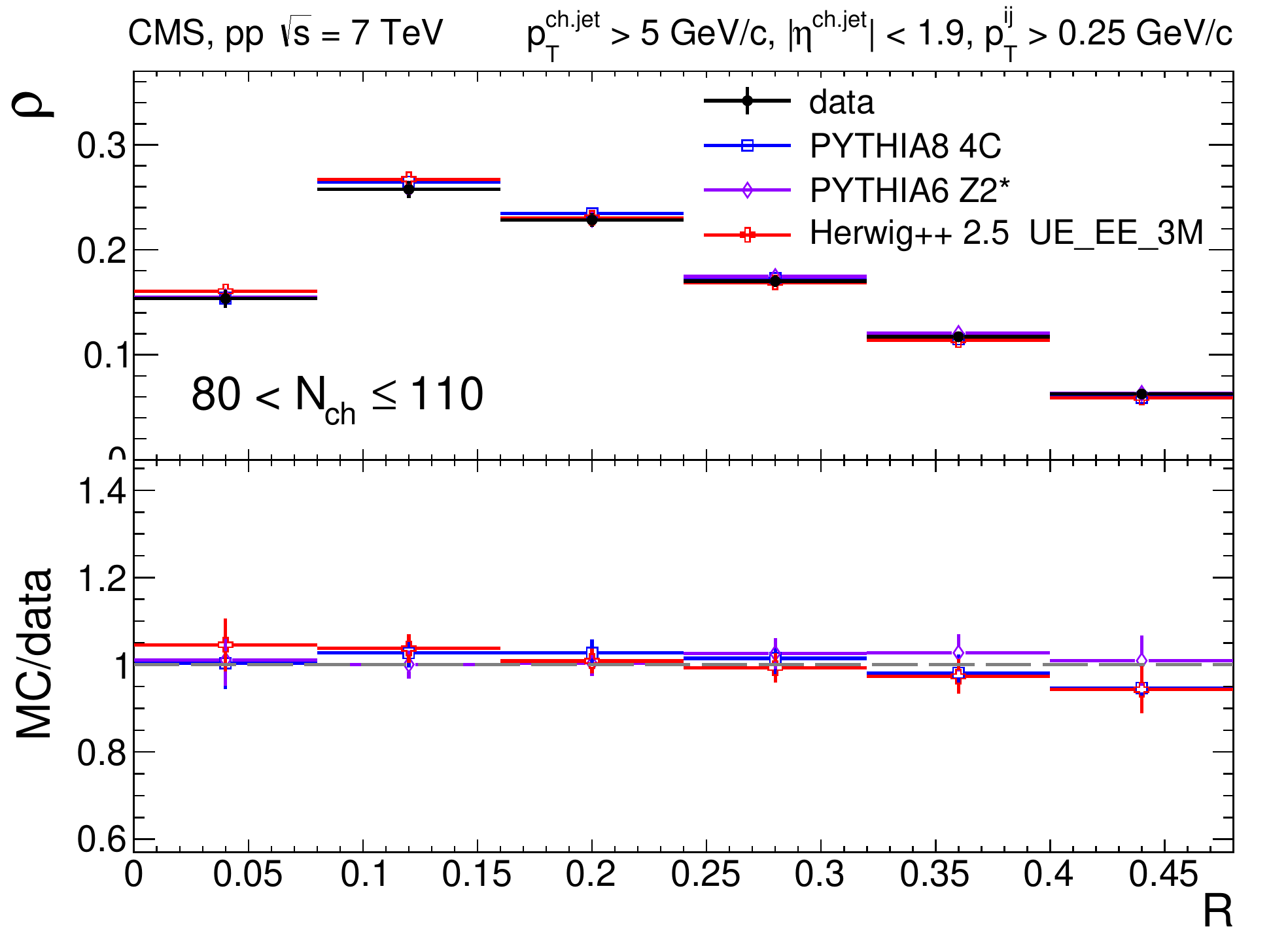}%
\caption{Normalized charged-particle jet $\PT$ density $\rho$ in ring zones as a function of distance to the jet axis $R$
  for events with $80<N_\text{ch}(\abs{\eta}<2.4) \leq 110$ measured in the data (solid dots) compared to
  various MC predictions (empty markers). Error bars denote the total uncertainties.}%
\label{jetstruct70100}%
\end{center}
\end{figure}

\begin{figure}[hbtp]
\begin{center}
\includegraphics[width=\cmsFigWidth]{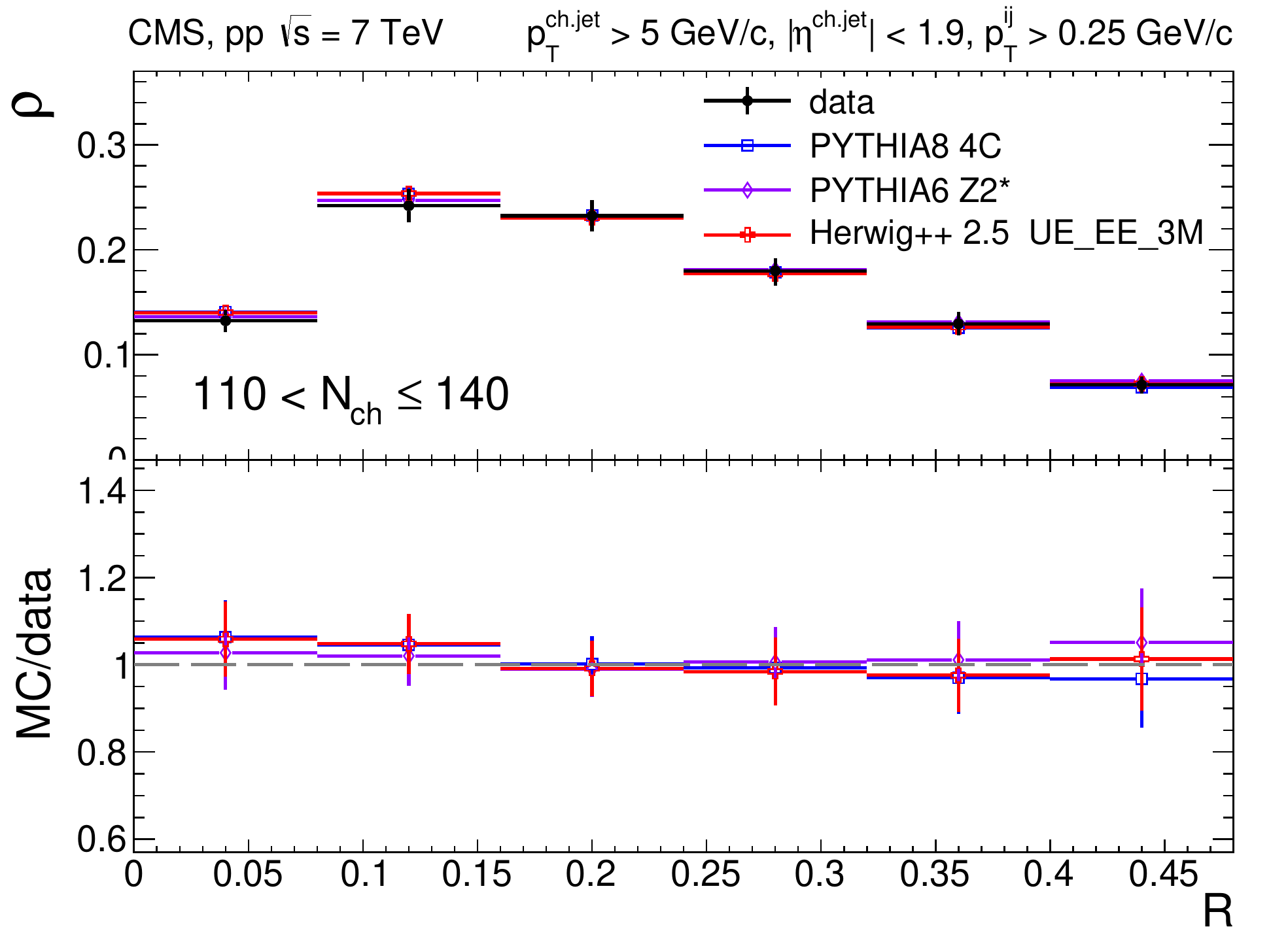}%
\caption{Normalized charged-particle jet $\PT$ density $\rho$ in ring zones as a function of distance to the jet axis $R$
  for events with $110<N_\text{ch}(\abs{\eta}<2.4) \leq 140$ measured in the data (solid dots) compared to
  various MC predictions (empty markers). Error bars denote the total uncertainties.}%
\label{jetstruct100130}%
\end{center}
\end{figure}

The jet width provides important information for characterizing the internal jet radiation dynamics. In this
analysis, we quantitatively study the jet width through the $\PT$ charged-particle density in ring zones with respect to the jet
center, defined as:
\begin{equation}
\rho = \left\langle \frac{1}{\PT^\text{ch. jet}}  \frac{\delta \PT^\text{ch. particles}}{\delta R}\right\rangle_\text{ch. jets},
\label{rhojets}
\end{equation}
where $R = \sqrt {(\phi -\phi_\text{jet})^2+(\eta -\eta _\text{jet})^2}$ is the distance of each charged particle from the jet
axis. Larger values of $\rho(R)$ denote a larger transverse momentum fraction in a particular annulus.
Jets with $\PT^\text{ch. jet}\geq5\GeVc$ are selected for the study. Data are compared with MC predictions
in five multiplicity intervals as shown in Figs.~\ref{jetstruct1030}--\ref{jetstruct100130}.
The dependencies shown in Figs.~\ref{jetstruct1030}--\ref{jetstruct100130} indicate that the jet width increases with $N_\text{ch}$,
which can be partly explained by the larger contribution of the UE to jets when $N_\text{ch}$ increases and partly
by softer, consequently larger-angle, hadronization, which follows from the intrinsic bias introduced by the
requirement of very large values of $N_\text{ch}$.
In low-multiplicity events, jets are narrower than predicted by \PYTHIA and {\HERWIG},
whereas in high-multiplicity events they are of comparable width as predicted by the MC event generators.
For events with  $10< N_\text{ch} \leq 50$, the \PYTHIA~8 model with MPI switched-off shows jet widths
that are close to the ones predicted by the models that include MPI,
but  it produces too hard jets, which are very collimated, in the bin $50< N_\text{ch} \leq 80$.
The patterns observed in the data show that the models need to be readjusted to reproduce the activity in the
innermost ring zone of the jet as a function of event multiplicity.

\section{Conclusions}\label{conclusion}

The characteristics of particle production in pp collisions at $\sqrt{s}=7$\TeV have been
presented as a function of the event charged-particle multiplicity ($N_\text{ch}$) by separating
the measured charged particles into those belonging to jets and those belonging to the underlying event.
Charged particles are measured within the pseudorapidity range $\abs{\eta}<2.4$ for transverse momenta $\PT>0.25\GeVc$
and charged-particle jets are reconstructed with $\PT~>5\GeVc$ with charged-particle information only.
The distributions of jet $\PT$, average $\PT$ of UE charged-particles and jets, jet rates, and jet shapes have
been studied as functions of $N_\text{ch}$ and compared to the predictions of the \PYTHIA and {\HERWIG} event
generators.

The average trends observed in the data are described by the QCD event generators but the
quantitative agreement, in particular at the highest multiplicity, is not as good. The mean transverse momentum
of inclusive and UE charged-particles and charged-jets, as well as the charged-jet rates, all rise
with $N_\text{ch}$ as expected for an increased fraction of (harder) multiple parton scatterings in more
central pp collisions resulting in increasingly higher multiplicity. On the other hand, the average $\PT$
of the intrajet constituents and the leading charged-particle of the jets decrease (logarithmically) with
increasing $N_\text{ch}$ as a result of a selection bias: final states with a larger number of hadrons result
from (mini)jets which fragment into more, and thus softer, hadrons. The characteristics of the highest
multiplicity pp events result from two seemingly opposite trends: a large number of parton interactions
with increasingly harder (mini)jets, combined with an overall softer distribution of final-state hadrons.

The detailed features of the $N_\text{ch}$-dependence of the jet and the UE properties differ from the MC
predictions. In general, \PYTHIA (and in particular \PYTHIA~6 tune Z2*) reproduces the data better than
{\HERWIG} for all observables measured. Of special interest is the large difference between the measured jet
$\PT$-differential spectra and the simulation predictions for the highest-multiplicity bins, above $N_\text{ch}$=80.
In these bins jets are softer, and less abundant than predicted by \PYTHIA, which explains the observed
larger event sphericity compared to predictions~\cite{AL}. The MC models also fail to fully describe the intrajet
spectra. The deviation of simulation predictions from the data for the spectra of the leading intrajet
particle is small in comparison to the variation between different models and their tunes, but systematic.
In low-multiplicity events, jets are narrower than predicted by \PYTHIA and {\HERWIG},
whereas in high-multiplicity events their widths are as predicted by the MC event generators.
At the same time, the characteristics of the UE are well reproduced by most of the MC event generators in all
the multiplicity bins considered.

The results obtained in this study are of importance both for improving the MC description of the data
and for getting a firmer grasp on the fundamental mechanisms of multi-particle production in hadronic collisions
at LHC energies. Current event generators tuned to reproduce the inelastic LHC data cannot describe
within a single approach the dependence of various quantities on event multiplicity. This is
especially true in the high-multiplicity range, where \PYTHIA produces many particles because of increased
high-$\PT$ jet contribution and {\HERWIG}++ seems to contain too many soft-parton scatterings.
The results of \PYTHIA with MPI switched off, demonstrate that the MPI mechanism is
critical for reproducing the measured properties of the jets and UE for moderate and large charged-particle
multiplicities. Taken together, the MC predictions globally bracket the data and indicate possible ways for
improving the parameter tuning and/or including new model ingredients.

\section*{Acknowledgments}
{\tolerance=800
We congratulate our colleagues in the CERN accelerator departments for the excellent performance
of the LHC and thank the technical and administrative staffs at CERN and at other CMS institutes
for their contributions to the success of the CMS effort. In addition, we gratefully acknowledge
the computing centres and personnel of the Worldwide LHC Computing Grid for delivering so
effectively the computing infrastructure essential to our analyses. Finally, we acknowledge the
enduring support for the construction and operation of the LHC and the CMS detector provided by
the following funding agencies: BMWF and FWF (Austria); FNRS and FWO (Belgium); CNPq, CAPES, FAPERJ,
and FAPESP (Brazil); MES (Bulgaria); CERN; CAS, MoST, and NSFC (China); COLCIENCIAS (Colombia);
MSES (Croatia); RPF (Cyprus); MoER, SF0690030s09 and ERDF (Estonia); Academy of Finland, MEC, and
HIP (Finland); CEA and CNRS/IN2P3 (France); BMBF, DFG, and HGF (Germany); GSRT (Greece); OTKA and
NKTH (Hungary); DAE and DST (India); IPM (Iran); SFI (Ireland); INFN (Italy); NRF and WCU (Republic of
Korea); LAS (Lithuania); CINVESTAV, CONACYT, SEP, and UASLP-FAI (Mexico); MSI (New Zealand);
PAEC (Pakistan); MSHE and NSC (Poland); FCT (Portugal); JINR (Dubna); MON, RosAtom, RAS and RFBR
(Russia); MESTD (Serbia); SEIDI and CPAN (Spain); Swiss Funding Agencies (Switzerland); NSC (Taipei);
ThEPCenter, IPST, STAR and NSTDA (Thailand); TUBITAK and TAEK (Turkey); NASU (Ukraine); STFC
(United Kingdom); DOE and NSF (USA).

Individuals have received support from the Marie-Curie programme and the European Research Council
and EPLANET (European Union); the Leventis Foundation; the A. P. Sloan Foundation; the Alexander von
Humboldt Foundation; the Belgian Federal Science Policy Office; the Fonds pour la Formation \'a la
Recherche dans l'Industrie et dans l'Agriculture (FRIA-Belgium); the Agentschap voor Innovatie door
Wetenschap en Technologie (IWT-Belgium); the Ministry of Education, Youth and Sports (MEYS) of
Czech Republic; the Council of Science and Industrial Research, India; the Compagnia di San Paolo
(Torino); the HOMING PLUS programme of Foundation for Polish Science, cofinanced by EU, Regional
Development Fund; and the Thalis and Aristeia programmes cofinanced by EU-ESF and the Greek NSRF.
\par
}

\bibliography{auto_generated}
\cleardoublepage \appendix\section{The CMS Collaboration \label{app:collab}}\begin{sloppypar}\hyphenpenalty=5000\widowpenalty=500\clubpenalty=5000\textbf{Yerevan Physics Institute,  Yerevan,  Armenia}\\*[0pt]
S.~Chatrchyan, V.~Khachatryan, A.M.~Sirunyan, A.~Tumasyan
\vskip\cmsinstskip
\textbf{Institut f\"{u}r Hochenergiephysik der OeAW,  Wien,  Austria}\\*[0pt]
W.~Adam, T.~Bergauer, M.~Dragicevic, J.~Er\"{o}, C.~Fabjan\cmsAuthorMark{1}, M.~Friedl, R.~Fr\"{u}hwirth\cmsAuthorMark{1}, V.M.~Ghete, N.~H\"{o}rmann, J.~Hrubec, M.~Jeitler\cmsAuthorMark{1}, W.~Kiesenhofer, V.~Kn\"{u}nz, M.~Krammer\cmsAuthorMark{1}, I.~Kr\"{a}tschmer, D.~Liko, I.~Mikulec, D.~Rabady\cmsAuthorMark{2}, B.~Rahbaran, C.~Rohringer, H.~Rohringer, R.~Sch\"{o}fbeck, J.~Strauss, A.~Taurok, W.~Treberer-Treberspurg, W.~Waltenberger, C.-E.~Wulz\cmsAuthorMark{1}
% ...
\vskip\cmsinstskip
% ...
1:~~Also at Vienna University of Technology, Vienna, Austria\\
2:~~Also at CERN, European Organization for Nuclear Research, Geneva, Switzerland\\
% ...

\end{sloppypar}
\end{document}